\documentclass[a4paper,12pt]{article}
\usepackage{bm,subeqnarray,eqsection,indent,cite}

\usepackage{cite,hyperref}
\usepackage{amsmath,amssymb,amsfonts}

\usepackage{graphicx}

\newcommand{\be}{\begin{equation}}
\newcommand{\ee}{\end{equation}}

\newcommand{\ef}[1]{\, #1}

\newcommand{\OSP}{\mathop{\rm OSP}\nolimits}
\newcommand{\osp}{\mathop{\rm osp}\nolimits}
\newcommand{\ssp}{\mathop{\rm sp}\nolimits}

\newcommand{\<}{\langle}
\renewcommand{\>}{\rangle}

\newcommand{\reff}[1]{(\ref{#1})}

\newcommand{\nv}{{\bf n}}
\def\psibar{{\bar{\psi}}}

\def\epsilonbar{{\bar{\epsilon}}}

\def\bt{{\bf t}}
\def\bv{{\bf v}}
\def\bw{{\bf w}}

\newcommand{\dx}[1] {d{#1}}
\def\ham{\mathcal{H}}

\newcommand{\sgn}{\mathop{\rm sgn}\nolimits}

\newtheorem{defin}{Definition}[section]

\newtheorem{proposition}[defin]{Proposition}

\newtheorem{lemma}[defin]{Lemma}

\newtheorem{theorem}[defin]{Theorem}

\newtheorem{corollary}[defin]{Corollary}

\renewcommand{\emptyset}{\varnothing}
\renewcommand{\setminus}{\smallsetminus}
\def\proof{\par\medskip\noindent{\sc Proof.\ }}

\def\firstproof{\par\medskip\noindent{\sc First proof.\ }}
\def\secondproof{\par\medskip\noindent{\sc Second proof.\ }}
\newcommand{\qed}{\quad $\Box$ \medskip \medskip}

\def\R{{\mathbb R}}
\def\C{{\mathbb C}}

\newcommand{\scrc}{{\mathcal{C}}}
\newcommand{\scrd}{{\mathcal{D}}}

\newcommand{\scrf}{{\mathcal{F}}}

\newcommand{\scro}{{\mathcal{O}}}

\newcommand{\scrt}{{\mathcal{T}}}

\newenvironment{scarray}{
          \textfont0=\scriptfont0
          \scriptfont0=\scriptscriptfont0
          \textfont1=\scriptfont1
          \scriptfont1=\scriptscriptfont1
          \textfont2=\scriptfont2
          \scriptfont2=\scriptscriptfont2
          \textfont3=\scriptfont3
          \scriptfont3=\scriptscriptfont3
        
        \begin{array}{c}}{\end{array}}

\begin{document}

\title{%
Grassmann Integral Representation \\ for Spanning Hyperforests}

\author{Sergio Caracciolo$^{1}$, Alan D. Sokal$^{2,3}$,
        Andrea Sportiello$^{1}$  \\[4mm]
\hspace*{-1cm}
{\small 
$^{1}$  Dipartimento di Fisica dell'Universit\`a degli Studi di Milano
        and INFN, Sezione di Milano,} \\[-1mm]
{\small via Celoria 16, I-20133 Milano, Italy}\\
\hspace*{-1cm}
{\small $^{2}$ Department of Physics, New York University,
      4 Washington Place, New York, NY 10003, USA}\\
\hspace*{-1cm}
{\small $^{3}$ Department of Mathematics,
           University College London, London WC1E 6BT, UK} \\[4mm]
\hspace*{-1cm}
{\tt \small 
S{}erg{}io.Cara{\phantom{\rule{0pt}{0pt}}}cci{}olo@mi.i{}nfn.it, 
so{}k{\phantom{\rule{0pt}{0pt}}}al@n{}yu.edu, 
Andr{}ea.Sportie{\phantom{\rule{0pt}{0pt}}}llo@mi.in{}fn.it}
   \\[1cm]
}

\date{June 10, 2007 \\[1mm] revised October 2, 2007}

\maketitle

\begin{abstract}
Given a hypergraph $G$, we introduce a Grassmann algebra over the
vertex set, and show that a class of Grassmann integrals permits
an expansion in terms of spanning hyperforests.
Special cases provide the generating functions for rooted and unrooted
spanning (hyper)forests and spanning (hyper)trees.
All these results are generalizations of Kirchhoff's matrix-tree theorem.
Furthermore, we show that the class of integrals describing
unrooted spanning (hyper)forests is induced by a theory with an
underlying $\OSP(1|2)$ supersymmetry.
\end{abstract}

PACS: 05.50.+q, 02.10.Ox, 11.10.Hi, 11.10.Kk

\medskip

Keywords: Graph, hypergraph, forest, hyperforest, matrix-tree theorem,
Grassmann algebra, Grassmann integral, $\sigma$-model, supersymmetry.

\newpage

\section{\label{sec:intro}Introduction}

Kirchhoff's matrix-tree theorem \cite{Kirchhoff,Tutte,Nerode}
and its generalizations \cite{Chaiken,Moon,abdes},
which express the generating polynomials of spanning trees
and rooted spanning forests in a graph
as determinants associated to the graph's Laplacian matrix,
play a central role in electrical circuit theory \cite{Balabanian_69,Chen}
and in certain exactly-soluble models in statistical mechanics
\cite{Duplantier_88,Wu_02}.

Like all determinants, those arising in Kirchhoff's theorem
can be rewritten as Gaussian integrals
over fermionic (Grassmann) variables.
Indeed, the use of Grassmann--Berezin calculus~\cite{Berezin}
has provided an interesting short-cut toward the classical matrix-tree result
as well as generalizations thereof~\cite{abdes,us_prl}.
For instance, Abdesselam~\cite{abdes} has obtained in a simple way
the recent pfaffian-tree theorem~\cite{Masbaum1,Masbaum2,Hirschman}
and has generalized it to a hyperpfaffian-cactus theorem.

In a recent letter \cite{us_prl}
we proved a far-reaching generalization of Kirchhoff's theorem,
in which a large class of combinatorial objects are represented
by suitable {\em non-Gaussian}\/ Grassmann integrals.
In particular, we showed how the
generating function of spanning forests in a graph,
which arises as the $q \to 0$ limit of the partition function of the
$q$-state Potts model \cite{Stephen_76,Wu_77,Jacobsen,alan},
can be represented as a Grassmann integral
involving a quadratic (Gaussian) term together with
a special nearest-neighbor four-fermion interaction.
Furthermore, this fermionic model possesses an $\OSP(1|2)$ supersymmetry.

This fermionic formulation is also well-suited to the use of
standard field-theoretic machinery.
For example, in \cite{us_prl} we obtained the renormalization-group flow
near the spanning-tree (free-field) fixed point for the spanning-forest
model on the square lattice, and in \cite{us_tri} this was extended
to the triangular lattice.

In the present paper we would like to extend
the fermionic representation of spanning forests from graphs to hypergraphs.
Hypergraphs are a generalization of graphs in which the edges
(now called hyperedges) can connect
more than two vertices~\cite{Diestel,berge1,berge2}.
In physics, hypergraphs arise quite naturally whenever one studies a
$k$-body interaction with $k>2$.\footnote{
   For examples in the recent physics literature where
   the hypergraph concept is used,
   see for instance~\cite{Grimmett_94,rwz,CSp,Castellani_03}.
}
We shall show here how the
generating function of spanning hyperforests in a hypergraph,
which arises as the $q \to 0$ limit of the partition function of the
$q$-state Potts model on the hypergraph \cite{Grimmett_94},
can be represented as a Grassmann integral
involving a quadratic term together with
special multi-fermion interactions associated to the hyperedges.
Once again, this fermionic model possesses an $\OSP(1|2)$ supersymmetry.
This extension from graphs to hypergraphs is thus not only natural,
but actually sheds light on the underlying supersymmetry.

Let us begin by recalling briefly the combinatorial identities
proven in \cite{us_prl}, which come in several levels of generality.
Let $G=(V,E)$ be a finite undirected graph with vertex set $V$
and edge set $E$.  To each edge $e$ we associate a weight $w_e$,
which can be a real or complex number or, more generally,
a formal algebraic variable;
we then define the {\em (weighted) Laplacian matrix}\/
$L = (L_{ij})_{i,j \in V}$ for the graph $G$ by
\begin{equation}
   L_{ij}  \;=\;  
   \begin{cases}
      -w_{ij}                        & \hbox{if }  i \neq j  \\
      \sum\limits_{k \neq i} w_{ik}  & \hbox{if }  i = j
   \end{cases}
\end{equation}
We introduce, at each vertex $i \in V$,
a pair of Grassmann variables $\psi_i$, $\psibar_i$,
which obey the usual rules for Grassmann integration \cite{Berezin,Zinn-Justin}.
Our identities show that certain Grassmann integrals over $\psi$ and $\psibar$
can be interpreted as generating functions
for certain classes of combinatorial objects on $G$.

Our most general identity concerns the operators $Q_\Gamma$
associated to arbitrary connected subgraphs
$\Gamma = (V_\Gamma, E_\Gamma)$ of $G$
via the formula
\begin{equation}
   Q_\Gamma  \;=\;
   \left( \prod_{e \in E_\Gamma} w_e \right)
   \left( \prod_{i \in V_\Gamma} \psibar_i \psi_i \right)
   \;.
\end{equation}
(Note that each $Q_\Gamma$ is even and hence commutes with
the entire Grassmann algebra.)
We prove the very general identity
\begin{equation}
  \int \! \scrd(\psi,\psibar)
     \; \exp\!\left[ \psibar L \psi + \sum\limits_\Gamma t_\Gamma Q_\Gamma
              \right]
   \;=
       \sum_{\begin{scarray}
                H \, \hbox{\scriptsize spanning} \subseteq G \\
                H = (H_1,\ldots,H_{\ell})
             \end{scarray}}
       \left( \prod_{e \in H} w_e \right)
       \prod_{\alpha=1}^{\ell} W(H_\alpha)
       \;,
 \label{eq.genfun2}
\end{equation}
where the sum runs over spanning subgraphs $H \subseteq G$
consisting of connected components $(H_1,\ldots,H_{\ell})$,
and the weights $W(H_\alpha)$ are defined by
\begin{equation}
   W(H_\alpha)  \;=\; \sum_{\Gamma \prec H_\alpha} t_\Gamma   \;,
\end{equation}
where $\Gamma \prec H_\alpha$ means that
$H_\alpha$ contains $\Gamma$ and contains no cycles other than those
lying entirely within $\Gamma$.

Let us now specialize \reff{eq.genfun2} to the case in which
$t_\Gamma = t_i$ when $\Gamma$ consists of a single vertex $i$ with no edges,
$t_\Gamma = u_e$ when $\Gamma$ consists of a pair of vertices $i,j$ linked by
an edge $e$, and $t_\Gamma = 0$ otherwise.
We then have
\begin{eqnarray}
   & &
   \!\!\!\!
     \int \! \scrd(\psi,\psibar)
        \, \exp\!\Big[
          \psibar L \psi \,
         +\, \sum\limits_i t_i \psibar_i \psi_i\,
         +\, \sum\limits_{\< ij \>} u_{ij} w_{ij}
                                    \psibar_i \psi_i \psibar_j \psi_j
               \Big]
      \nonumber \\
   & & \qquad\qquad
   \;=
       \sum_{\begin{scarray}
                F \in {\cal F}(G) \\
                F = (F_1,\ldots,F_{\ell})
             \end{scarray}}
       \!\!
       \Biggl( \prod_{e \in F} w_e \Biggr)
       \prod_{\alpha=1}^{\ell}  \Biggl( \sum_{i \in V(F_\alpha)} t_i \,+\,
                             \sum_{e \in E(F_\alpha)} u_e
                      \Biggr)
       \;,
   \qquad
 \label{eq.fourfermion.1}
\end{eqnarray}
where the sum runs over spanning forests $F$ in $G$
with components $F_1,\ldots,F_{\ell}$;
here $V(F_\alpha)$ and $E(F_\alpha)$ are, respectively,
the vertex and edge sets of the tree $F_\alpha$.

If we further specialize \reff{eq.fourfermion.1} to $u_e = -\lambda$
for all edges $e$ (where $\lambda$ is a global parameter), we obtain
\begin{eqnarray}
   & &
   \!\!\!\!
     \int \! \scrd(\psi,\psibar)
        \, \exp\!\Big[
          \psibar L \psi \,
         +\, \sum\limits_i t_i \psibar_i \psi_i\,
         -\, \lambda \sum\limits_{\< ij \>} w_{ij}
                                    \psibar_i \psi_i \psibar_j \psi_j
               \Big]
      \nonumber \\
   & & \qquad\qquad
   \;=
       \sum_{\begin{scarray}
                F \in {\cal F}(G) \\
                F = (F_1,\ldots,F_{\ell})
             \end{scarray}}
       \!\!
       \Biggl( \prod_{e \in F} w_e \Biggr)
       \prod_{\alpha=1}^{\ell}  \Biggl( \lambda \,+\,
                                   \sum_{i \in V(F_\alpha)} (t_i-\lambda)
                           \Biggr)
  \qquad
 \label{eq.fourfermion.2}
\end{eqnarray}
since $|E(F_\alpha)| = |V(F_\alpha)| - 1$.
If, in addition, we take $t_i = \lambda$ for all vertices $i$,
then we obtain
\begin{subeqnarray}
   & &
   \!\!\!\!
     \int \! \scrd(\psi,\psibar)
        \, \exp\!\Big[
          \psibar L \psi \,
         +\, \lambda \sum\limits_i \psibar_i \psi_i\,
         -\, \lambda \sum\limits_{\< ij \>} w_{ij}
                                    \psibar_i \psi_i \psibar_j \psi_j
               \Big]
      \nonumber \\
   & & \qquad\qquad
   \;=\!\!
       \sum_{F \in {\cal F}(G)}
       \Biggl( \prod_{e \in F} w_e \Biggr)
       \; \lambda^{k(F)}
           \qquad \\[1mm]
   & & \qquad\qquad
   \;=\;
      \lambda^{|V|} 
      \!\!\!\!
       \sum_{F \in {\cal F}(G)}
       \Biggl( \prod_{e \in F} \frac{w_e}{\lambda} \Biggr)
 \label{eq.fourfermion.3}
\end{subeqnarray}
where $k(F)$ is the number of component trees in the forest $F$;
this is the generating function of (unrooted) spanning forests of $G$.
Furthermore, as discussed in \cite{us_prl}
and in more detail in Section~\ref{sec:osp} below,
the model \reff{eq.fourfermion.3} possesses an $\OSP(1|2)$ invariance.
If, by contrast, in \reff{eq.fourfermion.2} we take $\lambda=0$
but $\{t_i\}$ general, we obtain
\begin{equation}
     \int \! \scrd(\psi,\psibar)
        \, \exp\!\Big[
          \psibar L \psi \,
         +\, \sum\limits_i t_i \psibar_i \psi_i\,
               \Big]
   \;=
       \sum_{\begin{scarray}
                F \in {\cal F}(G) \\
                F = (F_1,\ldots,F_{\ell})
             \end{scarray}}
       \!\!\!\!
       \Biggl( \prod_{e \in F} w_e \Biggr)
       \prod_{\alpha=1}^{\ell}  \Biggl( \sum_{i \in V(F_\alpha)} t_i \Biggr)
       \;,
 \label{eq.fourfermion.4}
\end{equation}
which is the formula representing {\em rooted}\/ spanning forests
(with a weight $t_i$ for each root $i$)
as a fermionic Gaussian integral (i.e., a determinant)
involving the Laplacian matrix
(this formula is a variant of the so-called
 ``principal-minors matrix-tree theorem'').

In this paper we shall not attempt to find the hypergraph analogue
of the general formula \reff{eq.genfun2},
but shall limit ourselves to finding analogues
of \reff{eq.fourfermion.1}--\reff{eq.fourfermion.4}.
The formulae to be presented here thus express the generating functions
of unrooted or rooted spanning hyperforests in a hypergraph
in terms of Grassmann integrals.
In particular, the hypergraph generalization of \reff{eq.fourfermion.3}
possesses the same $\OSP(1|2)$ supersymmetry that \reff{eq.fourfermion.3} does
(see Section~\ref{sec:osp}).

The proof given here of all these identities is purely algebraic
(and astonishingly simple);
the crucial ingredient is to recognize the role and the rules
of a certain Grassmann subalgebra (see Section~\ref{sec:grass}).
It turns out (Section~\ref{sec:osp})
that this subalgebra is nothing other than
the algebra of $\OSP(1|2)$-invariant functions,
though this is far from obvious at first sight.
The unusual properties of this subalgebra (see Lemma~\ref{lem.fAfB})
thus provide a deeper insight into the identities derived in \cite{us_prl}
as well as their generalizations to hypergraphs,
and indeed provide an alternate proof of
\reff{eq.fourfermion.2}--\reff{eq.fourfermion.4}.
Pictorially, we can say that it is the underlying supersymmetry that
is responsible for the cancellation of the cycles in the generating function,
leaving only those spanning (hyper)graphs that have no cycles,
namely, the (hyper)forests.

In particular, the limit of spanning hyper{\em trees}\/,
which is easily extracted from the general expression for
(rooted or unrooted) hyperforests,
corresponds in the $\OSP(1|2)$-invariant $\sigma$-model
to the limit in which the radius of the supersphere
tends to infinity,
so that the nonlinearity due to the curvature of the supersphere disappears.
However, the action is in general still non-quadratic,
so that the model is not exactly soluble.
(This is no accident: even the problem of determining whether
 there {\em exists}\/ a spanning hypertree in a given hypergraph
 is NP-complete \cite{Andersen_95}.)
Only in the special case of ordinary graphs is the action purely quadratic,
so that the partition function is given by a determinant,
corresponding to the statement of Kirchhoff's matrix-tree theorem.

The $\OSP(1|2)$-invariant fermionic models discussed in \cite{us_prl}
and the present paper can be written in three equivalent ways:
\begin{itemize}
   \item  As purely fermionic models,
       in which the supersymmetry is somewhat hidden.
   \item As $\sigma$-models with spins taking values in the unit supersphere
       in $\R^{1|2}$,
       in which the supersymmetry is manifest.
   \item As $N$-vector models [= $O(N)$-symmetric $\sigma$-models
       with spins taking values in the unit sphere of $\R^N$]
       analytically continued to $N=-1$.
\end{itemize}
The first two formulations (and their equivalence)
are discussed in Section~\ref{sec:osp} of the present paper.
Further aspects of this equivalence ---
notably, the role played by the Ising variables arising in
\reff{def.sigmai} and neglected here ---
will be discussed in more detail elsewhere \cite{us_forests_ON}.

In a subsequent paper \cite{us_Ward} we will discuss
the Ward identities associated to the $\OSP(1|2)$ supersymmetry,
and their relation to the combinatorial identities describing the
possible connection patterns among the (hyper)trees of a (hyper)forest.

%
%

The method proposed in the present paper has additional applications
not considered here.  With a small further effort,
a class of Grassmann integrals wider than
\reff{eq.Zgrass}/\reff{eq.corrfn.grass}
--- allowing products $\prod_{\alpha} f_{C_{\alpha}}^{(\lambda)}$
in the action
in place of the single operators $f_A^{(\lambda)}$ ---
can be handled.
Once again one obtains
a graphical expansion in terms of spanning hyperforests,
where now the weights have a more complicated dependence on the set of
hyperedges, thus permitting a description of certain natural
interaction patterns among the hyperedges of a hyperforest
(see Remark at the end of Section~\ref{sec:integrals}).
This extended model is, in fact, the {\em most general}\/ Hamiltonian
that is invariant under the $\OSP(1|2)$ supersymmetry.

\bigskip

The plan of this paper is as follows:
In Section~\ref{sec:graph} we recall the basic facts about
graphs and hypergraphs that will be needed in the sequel.
In Section~\ref{sec:potts} we define the $q$-state Potts model
on a hypergraph and prove the corresponding Fortuin--Kasteleyn representation.
(This section is unnecessary for the proof of the combinatorial identities
 that form the main focus of this paper, but it provides additional physical
 motivation.)
In Section~\ref{sec:grass} we introduce the Grassmann algebra
over the vertex set $V$, and study a subalgebra
with interesting and unusual properties,
which is generated by a particular family of even elements
$f_A^{(\lambda)}$ with $A \subseteq V$.
In Section~\ref{sec:integrals} we study a very general partition function
involving the operators $f_A^{(\lambda)}$, and we show how it can be
expressed as a generating function of spanning hyperforests in
a hypergraph with vertex set $V$.
In Section~\ref{sec:corrfn} we study a somewhat more general Grassmann
integral, which can be interpreted as a correlation function
in this same Grassmann model;
we show how it too can be expressed as a sum over spanning hyperforests.
In Section~\ref{sec:osp} we show that in one special case ---
namely, the hypergraph generalization of \reff{eq.fourfermion.3} ---
the model studied in the preceding sections 
can be rewritten as an $\OSP(1|2)$-invariant $\sigma$-model,
and indeed is the most general $\OSP(1|2)$-invariant Hamiltonian
involving $\psi$ and $\psibar$.
These $\sigma$-model formulae motivate the definition of $f_A^{(\lambda)}$
given in Section~\ref{sec:grass}, which might otherwise remain
totally mysterious.

In Appendix~\ref{app:grass.2} we prove a determinantal formula
for $f_A^{(\lambda)}$.
In Appendix~\ref{app:andrea} we present a graphical formalism
for proving both the classical matrix-tree theorem
and numerous extensions thereof,
which can serve as an alternative to the algebraic approach
used in the main body of this paper.

Let us stress that everything in this paper is mathematically rigorous,
with the possible exception of Section~\ref{sec:osp}.
Mathematicians unfamiliar with the Grassmann--Berezin calculus
can find a brief introduction in
\cite[Section~2]{abdes} or
\cite[Appendix~A]{CPSS_cayley}.

\section{Graphs and hypergraphs}
\label{sec:graph}

A (simple undirected finite) {\em graph}\/ is a pair $G=(V,E)$,
where $V$ is a finite set and $E$ is a collection (possibly empty)
of 2-element subsets of $V$.\footnote{
   To avoid notational ambiguities it should also be assumed that
   $E \cap V = \emptyset$.
   This stipulation is needed as protection against the mad set theorist
   who, when asked to produce a graph with vertex set $V = \{ 0,1,2 \}$,
   interprets this \`a la von Neumann as
   $V = \big\{ \, \emptyset ,\, \{\emptyset\} ,\, \{\emptyset,\{\emptyset\}\}
        \, \big\}$,
   so that the vertex~2 is indistinguishable from the edge $\{0,1\}$.
}
The elements of $V$ are the {\em vertices}\/ of the graph $G$,
and the elements of $E$ are the {\em edges}\/.
Usually, in a picture of a graph,
vertices are drawn as dots and edges as lines (or arcs).
Please note that, in the present definition,
loops
  (\setlength{\unitlength}{6pt}
   \begin{picture}(2.5,1)
   \put(0,0.5){\circle*{0.5}}
   \qbezier(0,0.5)(2,1.5)(2,0.5)
   \qbezier(0,0.5)(2,-0.5)(2,0.5)
   \end{picture})
and multiple edges
  (\setlength{\unitlength}{6pt}
   \begin{picture}(3.5,1)
   \put(0,0.5){\circle*{0.5}}
   \put(3,0.5){\circle*{0.5}}
   \qbezier(0,0.5)(1.5,1.5)(3,0.5)
   \qbezier(0,0.5)(1.5,-0.5)(3,0.5)
   \end{picture})
are not allowed.\footnote{
   This restriction is made mainly for notational simplicity.
   It would be easy conceptually to allow multiple edges,
   by defining $E$ as a {\em multiset}\/ (rather than a {\em set}\/)
   of 2-element subsets of $V$
   (cf.\ also footnote~\ref{note_multiset} below).
}
We write $|V|$ (resp.\ $|E|$) for the cardinality of the
vertex (resp.\ edge) set;
more generally, we write $|S|$ for the cardinality of any finite set $S$.

A graph $G'=(V',E')$ is said to be a {\em subgraph}\/ of $G$
(written $G'\subseteq G$)
in case $V'\subseteq V$ and $E'\subseteq E$.
If $V'=V$, the subgraph is said to be {\em spanning}\/.
We can, by a slight abuse of language, identify a spanning subgraph $(V,E')$
with its edge set $E'$.

A {\em walk}\/ (of length $k \ge 0$) connecting $v_0$ with $v_k$ in $G$
is a sequence
$(v_0, e_1, v_1, e_2, v_2,$ $\ldots,$ $e_k, v_k)$
such that all $v_i \in V$, all $e_i \in E$,
and $v_{i-1}, v_i \in e_i$ for $1 \le i \le k$.
A {\em path}\/ in $G$ is a walk in which
$v_0,\ldots,v_k$ are distinct vertices of $G$
and $e_1,\ldots,e_k$ are distinct edges of $G$.
A {\em cycle}\/ in $G$ is a walk in which
\begin{itemize}
   \item[(a)]  $v_0,\ldots,v_{k-1}$ are distinct vertices of $G$,
        and $v_k = v_0$
   \item[(b)]  $e_1,\ldots,e_k$ are distinct edges of $G$; and
   \item[(c)]  $k \ge 2$.\footnote{
       Actually, in a graph as we have defined it,
       all cycles have length $\ge 3$
       (because $e_1 \neq e_2$ and multiple edges are not allowed).
       We have presented the definition in this way
       with an eye to the corresponding definition for hypergraphs (see below),
       in which cycles of length 2 are possible.
}
\end{itemize}

The graph $G$ is said to be {\em connected}\/ if
every pair of vertices in $G$ can be connected by a walk.
The {\em connected components}\/ of $G$ are the
maximal connected subgraphs of $G$.
It is not hard to see that the property of being connected by a walk
is an equivalence relation on $V$,
and that the equivalence classes for this relation
are nothing other than the vertex sets of the connected components of $G$.
Furthermore, the connected components of $G$ are the induced subgraphs of $G$
on these vertex sets.\footnote{
   If $V' \subseteq V$, the {\em induced subgraph}\/ of $G$ on $V'$,
   denoted $G[V']$, is defined to be the graph $(V',E')$
   where $E'$ is the set of all the edges $e \in E$ that satisfy
   $e \subseteq V'$ (i.e., whose endpoints are both in $V'$).
}
We denote by $k(G)$ the number of connected components of $G$.
Thus, $k(G)=1$ if and only if $G$ is connected.

A {\em forest}\/ is a graph that contains no cycles.
A {\em tree}\/ is a connected forest.
(Thus, the connected components of a forest are trees.)
It is easy to prove, by induction on the number of edges, that
\begin{equation}
   |E| \,-\, |V| \,+\, k(G)  \;\ge\; 0
 \label{eq.eulerineq.graphs}
\end{equation}
for all graphs, with equality if and only if $G$ is a forest.
%

In a graph $G$, a {\em spanning forest}\/ (resp.\ {\em spanning tree}\/)
is simply a spanning subgraph that is a forest (resp.\ a tree).
We denote by $\scrf(G)$ [resp.\ $\scrt(G)$]
the set of spanning forests (resp.\ spanning trees) in $G$.
As mentioned earlier, we will frequently identify
a spanning forest or tree with its edge set.

Finally, we call a graph {\em unicyclic}\/ if it contains
precisely one cycle (modulo cyclic permutations and inversions
of the sequence $v_0, e_1, v_1, e_2, v_2, \ldots, e_k, v_k$).
It is easily seen that a connected unicyclic graph
consists of a single cycle together with trees
(possibly reduced to a single vertex) rooted at the vertices of the cycle.

Hypergraphs are the generalization of graphs
in which edges are allowed to contain more than two vertices.
Unfortunately, the terminology for hypergraphs varies substantially
from author to author,
so it is important to be precise about our own usage.
For us, a {\em hypergraph}\/ is a pair $G=(V,E)$,
where $V$ is a finite set and $E$ is a collection (possibly empty)
of subsets of $V$, each of cardinality $\ge 2$.\footnote{
   To avoid notational ambiguities it is assumed once again that
   $E \cap V = \emptyset$.
}
The elements of $V$ are the {\em vertices}\/ of the hypergraph $G$,
and the elements of $E$ are the {\em hyperedges}\/
(the prefix ``hyper'' can be omitted for brevity).
Note that we forbid hyperedges of 0 or 1 vertices
(some other authors allow these).\footnote{
   Our definition of hypergraph is the same as that of
   McCammond and Meier \cite{McCammond_04}.
   It is also the same as that of Grimmett \cite{Grimmett_94}
   and Gessel and Kalikow \cite{Gessel_05},
   except that these authors allow multiple edges and we do not:
   for them, $E$ is a {\em multiset}\/ of subsets of $V$
   (allowing repetitions), while for us
   $E$ is a {\em set}\/ of subsets of $V$ (forbidding repetitions).
   \protect\label{note_multiset}
}
We shall say that $A \in E$ is a {\em $k$-hyperedge}\/
if $A$ is a $k$-element subset of $V$.
A hypergraph is called {\em $k$-uniform}\/
if all its hyperedges are $k$-hyperedges.
Thus, a graph is nothing other than a 2-uniform hypergraph.

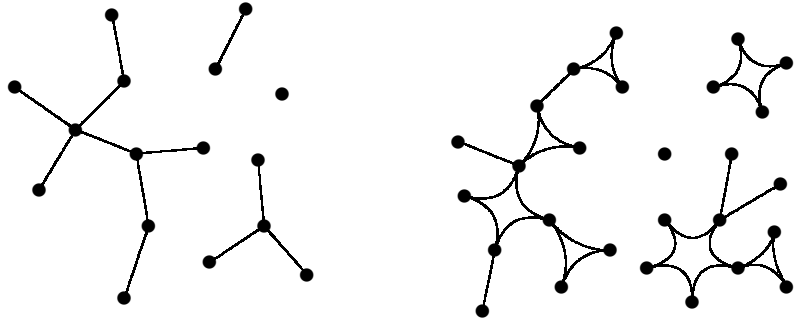
\begin{figure}[t]
\begin{center}
 \setlength{\unitlength}{8mm}
    \begin{picture}(5,5.3)
    \put(0.4,2.0){\circle*{0.2}}
    \put(1.0,3.0){\circle*{0.2}}
    \put(0.0,3.7){\circle*{0.2}}
    \put(1.8,3.8){\circle*{0.2}}
    \put(1.6,4.9){\circle*{0.2}}
    \put(2.0,2.6){\circle*{0.2}}
    \put(2.2,1.4){\circle*{0.2}}
    \put(1.8,0.2){\circle*{0.2}}
    \put(3.1,2.7){\circle*{0.2}}
    \put(3.3,4.0){\circle*{0.2}}
    \put(3.8,5.0){\circle*{0.2}}
    \put(4.4,3.6){\circle*{0.2}}
    \put(4.0,2.5){\circle*{0.2}}
    \put(4.1,1.4){\circle*{0.2}}
    \put(3.2,0.8){\circle*{0.2}}
    \put(4.8,0.6){\circle*{0.2}}
 \qbezier(0.4,2.0)(0.4,2.0)
 (1.0,3.0)
 \qbezier(1.0,3.0)(1.0,3.0)
 (0.0,3.7)
 \qbezier(1.0,3.0)(1.0,3.0)
 (1.8,3.8)
 \qbezier(1.8,3.8)(1.8,3.8)
 (1.6,4.9)
 \qbezier(1.0,3.0)(1.0,3.0)
 (2.0,2.6)
 \qbezier(2.0,2.6)(2.0,2.6)
 (2.2,1.4)
 \qbezier(2.2,1.4)(2.2,1.4)
 (1.8,0.2)
 \qbezier(2.0,2.6)(2.0,2.6)
 (3.1,2.7)
 \qbezier(3.3,4.0)(3.3,4.0)
 (3.8,5.0)
 \qbezier(4.1,1.4)(4.1,1.4)
 (4.0,2.5)
 \qbezier(4.1,1.4)(4.1,1.4)
 (3.2,0.8)
 \qbezier(4.1,1.4)(4.1,1.4)
 (4.8,0.6)
    \end{picture}
 \hspace*{15mm}
 \setlength{\unitlength}{8mm}
    \begin{picture}(5.5,5.3)
    \put(0.3,0){\circle*{0.2}}
    \put(0.5,1){\circle*{0.2}}
    \put(0,1.9){\circle*{0.2}}
    \put(0.9,2.4){\circle*{0.2}}
    \put(1.4,1.5){\circle*{0.2}}
    \put(1.6,0.4){\circle*{0.2}}
    \put(2.4,1){\circle*{0.2}}
    \put(1.9,2.7){\circle*{0.2}}
    \put(1.2,3.4){\circle*{0.2}}
    \put(1.8,4){\circle*{0.2}}
    \put(2.6,3.7){\circle*{0.2}}
    \put(2.5,4.6){\circle*{0.2}}
    \put(4.5,4.5){\circle*{0.2}}
    \put(4.1,3.7){\circle*{0.2}}
    \put(4.9,3.3){\circle*{0.2}}
    \put(5.3,4.1){\circle*{0.2}}
    \put(3.3,2.6){\circle*{0.2}}
    \put(4.4,2.6){\circle*{0.2}}
    \put(4.2,1.5){\circle*{0.2}}
    \put(3.3,1.5){\circle*{0.2}}
    \put(3,0.7){\circle*{0.2}}
    \put(3.75,0.15){\circle*{0.2}}
    \put(4.5,0.7){\circle*{0.2}}
    \put(5.3,0.4){\circle*{0.2}}
    \put(5.1,1.3){\circle*{0.2}}
 
    \put(5.2,2.1){\circle*{0.2}}
 \qbezier 
 (5.2,2.1)(5.2,2.1)
    (4.2,1.5)
    \put(-0.1,2.8){\circle*{0.2}}
 \qbezier (-0.1,2.8)
 (0.9,2.4)(0.9,2.4)
 
 \qbezier   (0.3,0)(0.5,1)(0.5,1)
 
 \qbezier   (0.5,1)(0.7,1.7)
    (0,1.9)
 \qbezier   (0,1.9)(0.7,1.7)
    (0.9,2.4)
 \qbezier   (0.9,2.4)(0.7,1.7)
    (1.4,1.5)
 \qbezier   (1.4,1.5)(0.7,1.7)
    (0.5,1)
 
 \qbezier   (1.4,1.5)(1.8,1)
    (1.6,0.4)
 \qbezier   (1.6,0.4)(1.8,1)
    (2.4,1)
 \qbezier   (2.4,1)(1.8,1)
    (1.4,1.5)
 
 \qbezier   (0.9,2.4)(1.3,2.8)
    (1.9,2.7)
 \qbezier   (1.9,2.7)(1.3,2.8)
    (1.2,3.4)
 \qbezier   (1.2,3.4)(1.3,2.8)
    (0.9,2.4)
 
 \qbezier
    (1.2,3.4)   (1.2,3.4)
    (1.8,4)
 
 \qbezier
    (1.8,4)(2.3,4.1)
    (2.6,3.7)
 \qbezier
    (2.6,3.7)(2.3,4.1)
    (2.5,4.6)
 \qbezier
    (2.5,4.6)(2.3,4.1)
    (1.8,4)
 
 \qbezier
    (4.5,4.5)(4.7,3.9)
    (4.1,3.7)
 \qbezier
    (4.1,3.7)(4.7,3.9)
    (4.9,3.3)
 \qbezier
    (4.9,3.3)(4.7,3.9)
    (5.3,4.1)
 \qbezier
    (5.3,4.1)(4.7,3.9)
    (4.5,4.5)
 
 \qbezier
    (4.4,2.6)   (4.4,2.6)
    (4.2,1.5)
 
 \qbezier
    (4.2,1.5)(3.75,0.91)
    (3.3,1.5)
 \qbezier
    (3.3,1.5)(3.75,0.91)
    (3,0.7)
 \qbezier
    (3,0.7)(3.75,0.91)
    (3.75,0.15)
 \qbezier
    (3.75,0.15)(3.75,0.91)
    (4.5,0.7)
 \qbezier
    (4.5,0.7)(3.75,0.91)
    (4.2,1.5)
 
 \qbezier
    (4.5,0.7)(5,0.9)
    (5.3,0.4)
 \qbezier
    (5.3,0.4)(5,0.9)
    (5.1,1.3)
 \qbezier
    (5.1,1.3)(5,0.9)
    (4.5,0.7)
    \end{picture}
\end{center}
\caption{A forest (left) and a hyperforest (right),
  each with four components. Hyperedges with more than two vertices are
  represented pictorially as star-like polygons.}
\label{fig.forest} 
\end{figure}

The definitions of subgraphs, walks, cycles, connected components,
trees, forests and unicyclics given above for graphs were explicitly chosen
in order to immediately generalize to hypergraphs:
it suffices to copy the definitions verbatim,
inserting the prefix ``hyper'' as necessary.
See Figure \ref{fig.forest} for examples of a forest and a hyperforest.

The analogue of the inequality \reff{eq.eulerineq.graphs} is the following:

\begin{proposition}
   \label{prop.eulerineq.hypergraphs}
Let $G=(V,E)$ be a hypergraph.  Then
\begin{equation}
   \sum_{A \in E} (|A|-1)  \,-\, |V|  \,+\,  k(G)  \;\ge\; 0
\ef,
  \label{eq.eulerineq.hypergraphs}
\end{equation}
with equality if and only if $G$ is a hyperforest.
\end{proposition}

\noindent
Proofs can be found, for instance,
in \cite[p.~392, Proposition~4]{berge1}
or \cite[pp.~278--279, Lemma]{Gessel_05}.

Please note one important difference between graphs and hypergraphs:
every connected graph has a spanning tree,
but not every connected hypergraph has a spanning hypertree.
Indeed, it follows from Proposition~\ref{prop.eulerineq.hypergraphs}
that if $G$ is a $k$-uniform connected hypergraph with $n$ vertices,
then $G$ can have a spanning hypertree only if $k-1$ divides $n-1$.
Of course, this is merely a necessary condition, not a sufficient one!
In fact, the problem of determining whether there exists
a spanning hypertree in a given connected hypergraph is NP-complete
(hence computationally difficult),
even when restricted to the following two classes of hypergraphs:
\begin{itemize}
   \item[(a)]  hypergraphs that are linear
       (each pair of edges intersect in at most one vertex)
       and regular of degree 3
       (each vertex belongs to exactly three hyperedges); or
   \item[(b)]  4-uniform hypergraphs containing a vertex which belongs to
       all hyperedges, and in which all other vertices have degree
       at most 3 (i.e., belong to at most three hyperedges)
\end{itemize}
(see \cite[Theorems~3 and 4]{Andersen_95}).
It seems to be an open question whether the problem remains NP-complete
for 3-uniform hypergraphs.

Finally, let us discuss how a connected hypergraph can be built up
one edge at a time.
Observe first that if $G=(V,E)$ is a hypergraph without isolated vertices,
then every vertex belongs to at least one edge
(that is what ``without isolated vertices'' means!),
so that $V = \bigcup\limits_{A \in E} A$.
In particular this holds if $G$ is a connected hypergraph
with at least two vertices.
So let $G=(V,E)$ be a connected hypergraph with $|V| \ge 2$;
let us then say that an ordering $(A_1,\ldots,A_m)$ of the hyperedge set $E$
is a {\em construction sequence}\/ in case all of the hypergraphs
$G_\ell = \big( \bigcup\limits_{i=1}^\ell A_i, \, \{A_1,\ldots,A_\ell\} \big)$
are connected ($1 \le \ell \le m$).
An equivalent condition is that
$\Big(\bigcup\limits_{i=1}^{\ell-1} A_i \Big) \cap A_\ell \neq \emptyset$
for $2 \le \ell \le m$.
We then have the following easy result:

\begin{proposition}
   \label{prop.construction.hypergraphs}
Let $G=(V,E)$ be a connected hypergraph with at least two vertices.
Then:
\begin{itemize}
   \item[(a)]  There exists at least one construction sequence.
   \item[(b)]  If $G$ is a hypertree, then for any construction sequence
      $(A_1,\ldots,A_m)$ we have
      $\Big| \Big(\bigcup\limits_{i=1}^{\ell-1} A_i \Big) \cap A_\ell \Big| = 1$
      for all $\ell$ $(2 \le \ell \le m)$.
   \item[(c)]  If $G$ is not a hypertree, then for any construction sequence
      $(A_1,\ldots,A_m)$ we have
      $\Big| \Big(\bigcup\limits_{i=1}^{\ell-1} A_i \Big) \cap A_\ell \Big|
           \ge 2$
      for at least one $\ell$.
\end{itemize}
\end{proposition}

\proof
(a)  The ``greedy algorithm'' works:
Let $A_1$ be {\em any}\/ hyperedge;  and at each stage $\ell \ge 2$,
let $A_\ell$ be {\em any}\/ hyperedge satisfying
$\Big(\bigcup\limits_{i=1}^{\ell-1} A_i \Big) \cap A_\ell \neq \emptyset$
(such a hyperedge has to exist, or else $G$ fails to be connected).

(b) and (c) are then easy consequences of
Proposition~\ref{prop.eulerineq.hypergraphs}.
\qed

\section{Potts model on a hypergraph}  \label{sec:potts} 

Let $q$ be a positive integer, and let $S$ be a set of cardinality $q$.
Then the {\em $q$-state Potts model}\/ on the hypergraph $G=(V,E)$
is defined as follows \cite{Grimmett_94}:
At each vertex $i \in V$ we place a {\em color}\/ (or {\em spin}\/)
variable $\sigma_i \in S$.
These variables interact via the Hamiltonian
\begin{equation}
\ham_{\rm Potts}(\sigma) \;=\; - \sum_{A \in E} J_A \delta_A(\sigma)
   \;,
\end{equation}
where $\{J_A\}_{A \in E}$ are a set of couplings associated
to the hyperedges of $G$,
and the Kronecker delta $\delta_A$ is defined for $A=\{i_1,\ldots,i_k\}$ by
\begin{equation}
\delta_A(\sigma)  \;=\;
\begin{cases}
1 & \hbox{if } \sigma_1= \cdots = \sigma_k  \\
0 & \hbox{otherwise }
\end{cases}
\end{equation}
The partition function $Z_G^{\rm Potts}$
is then the sum of $\exp[-\ham_{\rm Potts}(\sigma)]$
over all configurations $\sigma = \{ \sigma_i \}_{i \in V}$.

It is convenient to introduce the quantities $v_A = e^{J_A} - 1$;
we write ${\bf v} = \{ v_A \} _{A \in E}$ for the collection
of hyperedge weights.
We can then prove the Fortuin--Kasteleyn (FK) representation
\cite{Kasteleyn_69,Fortuin_72}
for the hypergraph Potts model \cite{Grimmett_94},
by following exactly the same method
as is used for graphs (see e.g.\ \cite[Section~2.2]{alan}):

\begin{proposition}[Fortuin--Kasteleyn representation]
Let $G=(V,E)$ be a hypergraph.  Then, for integer $q \ge 1$, we have
\begin{equation}
   Z_G^{\rm Potts}(q, {\bf v})
   \;\equiv\;
   \sum\limits_{\sigma \colon V \to S} \exp[-\ham_{\rm Potts}(\sigma)]
   \;=\;
   \sum\limits_{E' \subseteq E} q^{k(E')} \prod\limits_{A \in E'} v_A
   \;,
 \label{eq.FK.identity}
\end{equation}
where $k(E')$ denotes the number of connected components
in the hypergraph $(V,E')$.
\end{proposition}

\proof
We start by writing
\begin{equation}
   Z_G^{\rm Potts}(q, {\bf v})
   \;=\;
   \sum\limits_{\sigma \colon V \to S} \exp[-\ham_{\rm Potts}(\sigma)]
   \;=\;
   \sum\limits_{\sigma \colon V \to S}
     \prod\limits_{A \in E} \big[ 1 + v_A \delta_A (\sigma) \big]
   \;.
\end{equation}
Now expand out the product over $A \in E$,
and let $E' \subseteq E$ be the set of hyperedges for which the term
$v_A \delta_A(\sigma)$ is taken.
Now perform the sum over configurations $\{ \sigma_i \}_{i \in V}$:
in each connected component of the spanning subhypergraph $(V,E')$
the color $\sigma_i$ must be constant,
and there are no other constraints.
Therefore,
\begin{equation}
   Z_G^{\rm Potts}(q, {\bf v})
   \;=\;
   \sum\limits_{E' \subseteq E}  q^{k(E')}  \prod\limits_{A \in E'} v_A
   \;,
\end{equation}
as was to be proved.
\qed

Please note that the right-hand side of \reff{eq.FK.identity}
is a polynomial in $q$;
in particular, we can take it as the {\em definition}\/
of the Potts-model partition function $Z_G(q, {\bf v})$
for noninteger $q$.

Let us discuss in particular the various types of $q \to 0$ limits
that can be taken in the hypergraph Potts model,
by following a straightforward generalization of the method
that is used for graphs \cite[Section~2.3]{alan}:

The simplest limit is to take $q \to 0$ with fixed $\bv$.
{}From the definition \reff{eq.FK.identity} we see that
this selects out the spanning subhypergraphs $E' \subseteq E$
having the smallest possible number of connected components;
the minimum achievable value is of course $k(G)$ itself
(= 1 in case $G$ is connected, as it usually is).
We therefore have
\begin{equation}
   \lim_{q \to 0} q^{-k(G)} Z_G(q,\bv) \;=\;  C_G(\bv)
   \;,
\end{equation}
where
\begin{equation}
   C_G(\bv) \;=\; \!\!\! \sum\limits_{\begin{scarray}
                                       E' \subseteq E \\
                                       k(E') = k(G)
                                    \end{scarray}} \!\!
                       \prod_{A \in E'}  v_A
\end{equation}
is the generating polynomial of ``maximally connected spanning subhypergraphs''
(= {\em connected spanning subhypergraphs}\/ in case $G$ is connected).

A different limit can be obtained by taking $q \to 0$
with fixed values of $w_A = v_A/q^{|A|-1}$.
{}From \reff{eq.FK.identity} we have
\begin{equation}
   Z_G(q,\{ q^{|A|-1} w_A \})
   \;=\;
   \sum\limits_{E' \subseteq E} q^{k(E') + \sum\limits_{A \in E'} (|A|-1)}
        \prod\limits_{A \in E'} w_A
   \;.
\end{equation}
Using now Proposition~\ref{prop.eulerineq.hypergraphs},
we see that the limit $q \to 0$ selects out the spanning hyperforests:
\begin{equation}
   \lim_{q \to 0} q^{-|V|} Z_G(q,\{ q^{|A|-1} w_A \}) \;=\;  F_G(\bw)
   \;,
 \label{eq.limit.FG}
\end{equation}
where
\begin{equation}
   F_G(\bw) \;=\;  \sum\limits_{E' \in \scrf(G)}
                       \prod_{A \in E'}  w_A
   \label{def_F}
\end{equation}
is the generating polynomial of {\em spanning hyperforests}\/.

By a further limit we can obtain spanning hypertrees.
To see this, assume first that $G$ is connected
(otherwise there are no spanning hypertrees).
In $C_G(\bv)$, replace $v_A$ by $\lambda^{|A|-1} v_A$ and let $\lambda \to 0$;
then we pick out the connected spanning subhypergraphs having the
minimum value of $\sum_{A \in E'} (|A|-1)$,
which by Proposition~\ref{prop.eulerineq.hypergraphs}
are precisely the spanning hypertrees:
\begin{equation}
   \lim_{\lambda \to 0} \lambda^{-(|V|-1)} C_G(\{ \lambda^{|A|-1} v_A \})
   \;=\;
   T_G(\bv)
   \;,
 \label{eq.limit.TG1}
\end{equation}
where
\begin{equation}
   T_G(\bv) \;=\;  \sum\limits_{E' \in \scrt(G)} \prod_{A \in E'}  v_A
\end{equation}
is the generating polynomial of {\em spanning hypertrees}\/.
Alternatively, in $F_G(\bw)$,
replace $w_A$ by $\lambda^{|A|-1} w_A$ and let $\lambda \to \infty$;
then we pick out the spanning hyperforests having the
maximum value of $\sum_{A \in E'} (|A|-1)$,
which by Proposition~\ref{prop.eulerineq.hypergraphs}
are those with the minimum number of connected components,
i.e.\ again the spanning hypertrees:
\begin{equation}
   \lim_{\lambda \to \infty} \lambda^{-(|V|-1)} F_G(\{ \lambda^{|A|-1} w_A \})
   \;=\;
   T_G(\bw)
   \;.
 \label{eq.limit.TG2}
\end{equation}
There is, however, one important difference between the graph case
and the hypergraph case:
as discussed in Section~\ref{sec:graph},
every connected graph has a spanning tree,
but not every connected hypergraph has a spanning hypertree.
So the limits \reff{eq.limit.TG1} and \reff{eq.limit.TG2} can be zero.

%
%

\section{A Grassmann subalgebra with unusual properties}  \label{sec:grass} 

Let $V$ be a finite set of cardinality $n$.
For each $i \in V$ we introduce a pair $\psi_i$, $\psibar_i$
of generators of a Grassmann algebra
(with coefficients in $\R$ or $\C$).
We therefore have $2n$ generators,
and the Grassmann algebra (considered as a vector space over $\R$ or $\C$)
is of dimension $2^{2n}$.

For each subset $A \subseteq V$, we associate the monomial  
$\tau_A=\prod_{i \in A} \psibar_i \psi_i$, where $\tau_{\emptyset}=1$.
Please note that all these monomials are even elements of the
Grassmann algebra;
in particular, they commute with the whole Grassmann algebra.
Clearly, the elements $\{ \tau_A \}_{A \subseteq V}$
span a vector space of dimension $2^n$.
In fact, this vector space is a subalgebra, by virtue of the obvious relations
\begin{equation}
   \tau_A \, \tau_B
   \;=\;
   \begin{cases}
       \tau_{A \cup B}   & \hbox{if } A \cap B = \emptyset \\[2mm]
       0                 & \hbox{if } A \cap B \neq \emptyset
   \end{cases}
 \label{eq.tauAtauB}
\end{equation}

Let us now introduce another family of even elements of the Grassmann algebra,
also indexed by subsets of $V$,
which possesses very interesting and unusual properties.
For each subset $A \subseteq V$ and each number $\lambda$ (in $\R$ or $\C$),
we define the Grassmann element
\begin{equation}
f_A^{(\lambda)}   \; = \;
\lambda (1-|A|) \tau_A \,+\, \sum_{i \in A} \tau_{A \smallsetminus i}
  \,-\! \sum_{\begin{scarray}
                 i,j \in A \\
                 i \neq j
              \end{scarray}}
         \!
         \psibar_i \psi_j \tau_{A \smallsetminus \{i,j\}}
\ef.
  \label{eq.deff_A}
\end{equation}
(The motivation for this curious formula will be explained in
 Section~\ref{sec:osp}.)
For instance, we have
\begin{subeqnarray}
   f_\emptyset^{(\lambda)}
   & = &
   \lambda
         \\[2mm]
   f_{\{i\}}^{(\lambda)}
   & = &
   1  \qquad\hbox{for all } i
         \\[2mm]
   f_{\{i,j\}}^{(\lambda)} 
   & = &
   - \lambda \psibar_i\psi_i \psibar_j\psi_j
          \,+\, \psibar_i \psi_i \,+\, \psibar_j \psi_j
          \,-\, \psibar_i \psi_j \,-\, \psibar_j \psi_i
         \nonumber \\
   & = &
   - \lambda \psibar_i\psi_i \psibar_j\psi_j
          \,+\, (\psibar_i - \psibar_j) (\psi_i - \psi_j)
 \label{eq4.3}
\end{subeqnarray}
and in general
\begin{equation}
   f_{\{i_1,\ldots,i_k\}}^{(\lambda)}
   \;=\;
   \lambda (1-k) \tau_{\{i_1,\ldots,i_k\}}
     \,+\,  \sum_{\alpha=1}^k \tau_{\{i_1,\ldots,\not{i_\alpha},\ldots,i_k\}}
     \,-\, \!\!\! \sum_{\begin{scarray}
                     1 \le \alpha,\beta \le k \\
                     \alpha \neq \beta
                  \end{scarray}}
\!\!\!
           \psibar_{i_\alpha} \psi_{i_\beta}
           \tau_{\{i_1,\ldots,\not{i_\alpha},\ldots,\not{i_\beta},\ldots,i_k\}}
   \;.
\end{equation}
(Whenever we write a set $\{i_1,\ldots,i_k\}$,
 it is implicitly understood that the elements $i_1,\ldots,i_k$
 are all distinct.)
Clearly, each $f_A^{(\lambda)}$ is an even element in the Grassmann algebra,
and in particular it commutes with all the other elements of the
Grassmann algebra.

The definition \reff{eq.deff_A} can also be rewritten as
  \begin{equation}
  f_A^{(\lambda)}   \;=\;
       \biggl( \lambda (1-|A|) +
               \sum\limits_{i,j \in A} \partial_i \bar{\partial}_j
       \biggr)  \tau_A  \;=\;
       \Bigl( \lambda (1-|A|) +
               \partial \bar{\partial}
       \Bigr) \tau_A 
       \label{eq:new_A}
  \end{equation}
where $\partial_i = \partial/\partial \psi_i$
and $\bar{\partial}_i = \partial/\partial \psibar_i$
are the traditional anticommuting differential operators satisfying
$\partial_i \psi_j = \delta_{ij}$, $\partial_i \psibar_j = 0$,
$\bar{\partial}_i \psibar_j = \delta_{ij}$, $\bar{\partial}_i \psi_j = 0$
and the (anti-)Leibniz rule,
while $\partial = \sum\limits_{i \in V} \partial_{i}$
and $\bar{\partial} = \sum\limits_{i \in V} \bar{\partial}_{i}$.

Let us observe that
\begin{equation}
   f_A^{(\lambda)} \, \tau_B
   \;=\;
   \begin{cases}
       \tau_{A \cup B}   & \hbox{if } |A \cap B| = 1    \\[2mm]
       0                 & \hbox{if } |A \cap B| \ge 2
   \end{cases}
 \label{eq.fAtauB}
\end{equation}
as an immediate consequence of \reff{eq.tauAtauB}
[when $A \cap B = \{k\}$, only the second term in \reff{eq.deff_A}
 with $i=k$ survives].
Note, finally, the obvious relations
\begin{equation}
   \lim_{\lambda \to \infty} \frac{1}{\lambda} f_A^{(\lambda)}
   \;=\;
   (1-|A|) \tau_A
\end{equation}
and
\begin{equation}
   f_A^{(\lambda)} - f_A^{(\lambda')}
   \;=\;
   (\lambda-\lambda') (1-|A|) \tau_A   \;.
\end{equation}

We are interested in the subalgebra of the Grassmann algebra
that is generated by the elements $f_A^{(\lambda)}$
as $A$ ranges over all nonempty subsets of $V$,
for an arbitrary fixed value of $\lambda$.\footnote{
   One can also consider the smaller subalgebras
   generated by the elements $f_A^{(\lambda)}$
   as $A$ ranges over some collection ${\cal S}$ of subsets of $V$.
}
The key to understanding this subalgebra
is the following amazing identity:

\begin{lemma}
  \label{lem.fAfB}
Let $A, B \subseteq V$ with $A \cap B \neq \emptyset$.
Then
\begin{equation}
   f_A^{(\lambda)} \,f_B^{(\lambda)}
   \;=\;
   \begin{cases}
       f_{A \cup B}^{(\lambda)} & \textrm{if } |A \cap B| = 1 \\[2mm]
       0                        & \textrm{if } |A \cap B| \geq 2
   \end{cases}
 \label{eq.fafb}
\end{equation}
More generally,
\begin{equation}
   f_A^{(\lambda)} \,f_B^{(\lambda')}
   \;=\;
   \begin{cases}
       f_{A \cup B}^{(\lambda'')} & \textrm{if } |A \cap B| = 1 \\[2mm]
       0                        & \textrm{if } |A \cap B| \geq 2
   \end{cases}
 \label{eq.fafb.extended}
\end{equation}
where $\lambda''$ is the weighted average
\begin{equation}
   \lambda''
   \;=\;
   \frac{ (|A|-1) \lambda + (|B|-1) \lambda'}{|A|+|B|-2}
   \;=\;
   \frac{ (|A|-1) \lambda + (|B|-1) \lambda'}{|A \cup B| - 1}
   \;.
\end{equation}
\end{lemma}

\firstproof
The formula \reff{eq.fafb.extended} can be proven by
a direct (but lengthy) calculation within the Grassmann algebra
that makes explicit a sort of fermionic-bosonic cancellation.
Details can be found in the first preprint version of this article
({\tt arXiv:0706.1509v1});  see especially footnote 10 there.
\qed

\secondproof
We are grateful to an anonymous referee for suggesting the following
simple and elegant proof using the differential operators
$\partial$ and $\bar{\partial}$:

Since $\partial^2 = \bar{\partial}^2=0$, we have
\begin{equation} 
\left( \partial \bar{\partial} \tau_A \right)
\left( \partial \bar{\partial} \tau_B  \right)
    \;=\;
\partial \bar{\partial} \left(  \tau_A \partial \bar{\partial} \tau_B  \right)
    \;=\;
\partial \bar{\partial} \left(  \tau_B \partial \bar{\partial} \tau_A  \right)
   \;,
\end{equation}
so that
\begin{eqnarray}
 f_A^{(\lambda)} \,f_B^{(\lambda')}
   & = &
 \lambda (1-|A|) \tau_A \partial \bar{\partial} \tau_B   \,+\,
 \lambda' (1-|B|) \tau_B \partial \bar{\partial} \tau_A
    \nonumber  \\
  & & \quad+\,\lambda  \lambda' (1-|A|) (1-|B|) \tau_A  \tau_B  \,+\,
   \partial \bar{\partial}
       \left(  \tau_A \partial \bar{\partial} \tau_B  \right)
   \;.
  \label{eq:lem}
\end{eqnarray}
If $|A \cap B|\geq 1$, then $\tau_A \tau_B=0$ and
\begin{equation}
  \tau_A \partial \bar{\partial} \tau_B 
  \;=\;
  \tau_B \partial \bar{\partial} \tau_A 
  \;=\;
  \begin{cases}
        \tau_{A \cup B}  & \textrm{if $|A \cap B| = 1$}  \\[2mm]
        0                & \textrm{if $|A \cap B| \ge 2$}
   \end{cases}
\end{equation}
This proves \reff{eq.fafb.extended}.
\qed

As a first consequence of Lemma~\ref{lem.fAfB}, we have:

\begin{corollary}
\label{corol:nil}
Let $A \subseteq V$ with $|A| \ge 2$. 
Then the Grassmann element $f_A^{(\lambda)}$ is nilpotent of order 2, i.e.
$$\big(f_A^{(\lambda)} \big)^2=0
\ef.$$ 
\end{corollary}

In particular, a product $\prod_{i=1}^m f_{A_i}^{(\lambda)}$
vanishes whenever there are any repetitions among the $A_1,\ldots,A_m$.
So we can henceforth concern ourselves with the case in which
there are no repetitions;  then $E = \{ A_1,\ldots,A_m \}$
is a {\em set}\/ (as opposed to a {\em multiset}\/)
and $G=(V,E)$ is a hypergraph.

By iterating Lemma~\ref{lem.fAfB}
and using Proposition~\ref{prop.construction.hypergraphs}, we easily obtain:

\begin{corollary}
  \label{corol:nil2}
Let $G=(V,E)$ be a \emph{connected} hypergraph.  Then
\begin{equation}
   \prod_{A\in E}  f_A^{(\lambda)}
   \;=\;
   \begin{cases}
        f_V^{(\lambda)}  & \textrm{if $G$ is a hypertree}  \\[2mm]
        0                & \textrm{if $G$ is not a hypertree}
   \end{cases}
\end{equation}
More generally,
\begin{equation}
   \prod_{A\in E}  f_A^{(\lambda_A)}
   \;=\;
   \begin{cases}
        f_V^{(\lambda_\star)}  & \textrm{if $G$ is a hypertree}  \\[2mm]
        0                & \textrm{if $G$ is not a hypertree}
   \end{cases}
\end{equation}
where $\lambda_\star$ is the weighted average
\begin{equation}
   \lambda_\star
   \;=\;
   \frac{ \, \sum\limits_{A \in E}  (|A|-1) \lambda_A \,
     }{
     \sum\limits_{A \in E}  (|A|-1)
   }
   \;=\;
   \frac{ \, \sum\limits_{A \in E}  (|A|-1) \lambda_A \,
     }{
\rule{0pt}{11.5pt}%
\bigl| \bigcup\limits_{A \in E} A \bigr| - 1
   }
   \;\,.
 \label{eq.weighted_lambda}
\end{equation}
\end{corollary}


We are now ready to consider the subalgebra of the Grassmann algebra
that is generated by the elements $f_A^{(\lambda)}$
as $A$ ranges over all nonempty subsets of $V$.
Recall first that a {\em partition}\/ of $V$
is a collection $\scrc = \{ C_{\gamma} \}$
of disjoint nonempty subsets $C_{\gamma} \subseteq V$
that together cover $V$.
We denote by $\Pi(V)$ the set of partitions of $V$.
If $V$ has cardinality $n$, then $\Pi(V)$ has cardinality $B(n)$,
the $n$-th \emph{Bell number} \cite[pp.~33--34]{Stanley_86}.
We remark that $B(n)$ grows asymptotically roughly like $n!$
\cite[Sections 6.1--6.3]{deBruijn_61}.

The following corollary specifies the most general product
of factors $f_A^{(\lambda)}$.
Of course, there is no need to consider sets $A$ of cardinality 1,
since $f_{\{i\}}^{(\lambda)} = 1$.
 
\begin{corollary}
  \label{corol:for}
Let $E$ be a collection (possibly empty) of subsets of $V$,
each of cardinality $\ge 2$.
\begin{itemize}
   \item[(a)]  If the hypergraph $G = (V,E)$ is a hyperforest,
        and $\{ C_{\gamma} \}$ is the partition of $V$
        induced by the decomposition of $G$ into connected components,
        then $\prod\limits_{A \in E} f_{A}^{(\lambda)} =
              \prod\limits_{\gamma} f_{C_{\gamma}}^{(\lambda)}$.
        More generally,
             $\prod\limits_{A \in E} f_{A}^{(\lambda_A)} =
              \prod\limits_{\gamma} f_{C_{\gamma}}^{(\lambda_\gamma)}$,
        where $\lambda_\gamma$ is the weighted average
        \reff{eq.weighted_lambda} taken over the hyperedges contained
        in $C_\gamma$.
   \item[(b)]  If the hypergraph $G = (V,E)$ is not a hyperforest,
        then $\prod\limits_{A \in E} f_{A}^{(\lambda)} = 0$,
        and more generally $\prod\limits_{A \in E} f_{A}^{(\lambda_A)} = 0$.
\end{itemize}
\end{corollary}

\proof
It suffices to apply Corollary~\ref{corol:nil2}
separately in each set $C_\gamma$,
where $\{ C_{\gamma} \}$ is the partition of $V$
induced by the decomposition of $G$ into connected components.
\qed

It follows from Corollary~\ref{corol:for} that
any polynomial (or power series) in the $\{ f_A^{(\lambda)}\}$
can be written as a linear combination of the quantities
$f_\scrc^{(\lambda)} = \prod_{\gamma} f_{C_{\gamma}}^{(\lambda)}$
for partitions $\scrc = \{ C_\gamma \} \in \Pi(V)$.


Using the foregoing results, we can simplify the Boltzmann weight
associated to a Hamiltonian of the form
\begin{equation}
\label{eq.hamlin}
   \mathcal{H}
   \;=\;
   - \sum_{A \in E} w_A f_A^{(\lambda)} 
   \;.
\end{equation}

\begin{corollary}
  \label{cor.exp}
Let $G=(V,E)$ be a hypergraph
(that is, $E$ is a collection of subsets of $V$, each of cardinality $\ge 2$).
Then
\begin{equation}
   \exp \Biggl( \sum_{A \in E} w_A f_A^{(\lambda)} \Biggr)
   \;= \!\!
   \sum_{\begin{scarray}
              F\in \scrf(G)  \\
              F = (F_1,\ldots,F_{\ell})
         \end{scarray}}
   \!\!
     \Biggl( \prod\limits_{A \in F} w_A \Biggr)
     \,
     \prod_{\alpha=1}^{\ell} f_{V(F_\alpha)}^{(\lambda)}
\ef,
 \label{eq.cor.exp}
\end{equation}
where the sum runs over spanning hyperforests $F$ in $G$
with components $F_1,\ldots, F_{\ell}$,
and $V(F_\alpha)$ is the vertex set of the hypertree $F_\alpha$.
More generally,
\begin{equation}
   \exp \Biggl( \sum_{A \in E} w_A f_A^{(\lambda_A)} \Biggr)
   \;= \!\!
   \sum_{\begin{scarray}
              F\in \scrf(G)  \\
              F = (F_1,\ldots, F_{\ell})
         \end{scarray}}
   \!\!
     \Biggl( \prod\limits_{A \in F} w_A \Biggr)
     \,
     \prod_{\alpha=1}^{\ell} f_{V(F_\alpha)}^{(\lambda_\alpha)}
\ef,
 \label{eq.cor.exp.bis}
\end{equation}
where $\lambda_\alpha$ is the weighted average \reff{eq.weighted_lambda}
taken over the hyperedges contained in the hypertree $F_\alpha$.
\end{corollary}

\proof
Since the $f_A^{(\lambda_A)}$ are nilpotent of order 2 and commuting,
we have
\begin{subeqnarray}
   \exp \Biggl( \sum_{A \in E} w_A f_A^{(\lambda_A)} \Biggr)
   & = &
   \prod\limits_{A\in E} \Big( 1 +  w_A f_A^{(\lambda_A)} \Big)
        \\[2mm]
   & = &
   \sum\limits_{E' \subseteq E}
      \Biggl( \prod_{A \in E'} w_A \Biggr)
      \Biggl( \prod_{A \in E'} f_A^{(\lambda_A)} \Biggr)
   \;.
\end{subeqnarray}
Using now Corollary~\ref{corol:for},
we see that the contribution is nonzero only when $(V,E')$ is a hyperforest,
and we obtain \reff{eq.cor.exp}/\reff{eq.cor.exp.bis}.
\qed

In a separate paper \cite{CSS:subalgebra}
we shall study in more detail the Grassmann subalgebra
that is generated by the elements $f_A^{(\lambda)}$
as $A$ ranges over all nonempty subsets of $V$.
In the present section we have seen that any element of this subalgebra
can be written as a linear combination of the quantities
$f_\scrc^{(\lambda)} = \prod_{\gamma} f_{C_{\gamma}}^{(\lambda)}$
for partitions $\scrc = \{ C_\gamma \} \in \Pi(V)$.
It turns out that the quantities
$f_\scrc^{(\lambda)}$ are linearly {\em dependent}\/
(i.e., an overcomplete set) as soon as $|V| \ge 4$.
We shall show \cite{CSS:subalgebra}, in fact, that
a vector-space basis for the subalgebra in question
is given by the quantities $f_\scrc^{(\lambda)}$
as $\scrc$ ranges over all {\em non-crossing}\/ partitions of $V$
(relative to any fixed total ordering of $V$).
It follows that the vector-space dimension of this subalgebra
is given by the {\em Catalan number}\/
$C_n = \frac{1}{n+1} \binom{2n}{n}$, where $n = |V|$.
This is vastly smaller than the Bell number $B(n)$,
which is the dimension that the subspace would have
if the $f_\scrc^{(\lambda)}$ were linearly independent.
[Indeed, one can see immediately that the $\{f_\scrc^{(\lambda)}\}$
 must be linearly dependent for all sufficiently large $n$,
 simply because the entire Grassmann algebra
 has dimension only $4^n \ll B(n)$.]
It also turns out \cite{CSS:subalgebra}
that all the relations among the $\{f_\scrc^{(\lambda)}\}$
are generated (as an ideal) by the elementary relations $R_{abcd} = 0$, where
\begin{multline}
   R_{abcd}  \;=\;
   \lambda f_{\{a,b,c,d\}}^{(\lambda)}
   \,-\, f_{\{b,c,d\}}^{(\lambda)}
   \,-\, f_{\{a,c,d\}}^{(\lambda)}
   \,-\, f_{\{a,b,d\}}^{(\lambda)}
   \,-\, f_{\{a,b,c\}}^{(\lambda)}
\\
\,+\, f_{\{a,b\}}^{(\lambda)} f_{\{c,d\}}^{(\lambda)}
\,+\, f_{\{a,c\}}^{(\lambda)} f_{\{b,d\}}^{(\lambda)}
\,+\, f_{\{a,d\}}^{(\lambda)} f_{\{b,c\}}^{(\lambda)}
\end{multline}
and $a,b,c,d$ are distinct vertices.


\section{Grassmann integrals for counting spanning hyperforests}
   \label{sec:integrals}

For any subset $A \subseteq V$ and any vector $\bt = (t_i)_{i \in V}$
of vertex weights, let us define the integration measure
\begin{equation}
 \label{intm}
  \mathcal{D}_{A,\bt}(\psi, \psibar)
  \; := \;
  \prod_{i \in A} \dx{\psi_i} \, \dx{\psibar_i} \, e^{ t_i \psibar_i \psi_i}
\ef.
\end{equation}
Our principal goal in this section is to provide a combinatorial
interpretation, in terms of spanning hyperforests, for the general
Grassmann integral (``partition function'')
\begin{subeqnarray}
   Z
   & = &
   \int \mathcal{D}(\psi, \psibar) \,
      \exp \biggl[ \, \sum_i t_i \psibar_i \psi_i
                      \,+\,  \sum_{A \in E} w_A f_A^{(\lambda)} \biggr]
         \\[2mm]
   & = &
   \int \mathcal{D}_{V,\bt}(\psi, \psibar) \,
      \exp \biggl[ \, \sum_{A \in E} w_A f_A^{(\lambda)} \biggr]
   \;,
 \label{eq.Zgrass}
\end{subeqnarray}
where $G=(V,E)$ is an arbitrary hypergraph
(that is, $E$ is an arbitrary collection of subsets of $V$,
 each of cardinality $\ge 2$)
and the $\{ w_A \} _{A \in E}$ are arbitrary hyperedge weights.
We also handle the slight generalization in which a separate parameter
$\lambda_A$ is used for each hyperedge $A$.


Our basic results are valid for an arbitrary vector $\bt = (t_i)_{i \in V}$
of ``mass terms''.  However, as we shall see, the formulae simplify
notably if we specialize to the case in which $t_i = \lambda$
for all $i \in V$.  This is not an accident, as it corresponds to the case
in which the action is $\OSP(1|2)$-invariant
(see Section~\ref{sec:osp}).

We begin with some formulae that allow us to integrate over the pairs
of variables $\psi_i, \psibar_i$ one at a time:

\begin{lemma}
  \label{lemma.inti}
Let $A \subseteq V$ and $i \in V$.  Then:
\begin{itemize}
 \item[(a)]
   ${\displaystyle \int} \dx{\psi_i} \, \dx{\psibar_i} \,
         e^{ t_i \psibar_i \psi_i} \, \tau_A
    \;=\;
    {\displaystyle
        \begin{cases}
            \tau_{A \smallsetminus i}  & \textrm{if }  i \in A   \\[2mm]
            t_i \tau_A                 & \textrm{if }  i \notin A
        \end{cases}
    }
   $
 \item[(b)]
   ${\displaystyle \int} \dx{\psi_i} \, \dx{\psibar_i} \,
         e^{ t_i \psibar_i \psi_i} \, f_A^{(\lambda)}
    \;=\;
    {\displaystyle
        \begin{cases}
            f_{A \smallsetminus i}^{(\lambda)}
                \,+\, (t_i-\lambda) \tau_{A \smallsetminus i}
                                       & \textrm{if }  i \in A   \\[2mm]
            t_i f_A^{(\lambda)}        & \textrm{if }  i \notin A
        \end{cases}
    }
   $
\end{itemize}
\end{lemma}

\proof
(a) is obvious, as is (b) when $i \notin A$.
To prove (b) when $i \in A$, we write
\begin{equation}
f_A^{(\lambda)}   \; = \;
\lambda (1-|A|) \tau_A \,+\, \sum_{j \in A} \tau_{A \smallsetminus j}
  \,-\! \sum_{\begin{scarray}
                 j,k \in A \\
                 j \neq k
              \end{scarray}}
         \!
         \psibar_j \psi_k \tau_{A \smallsetminus \{j,k\}}
\end{equation}
and integrate with respect to
$\dx{\psi_i} \, \dx{\psibar_i} \, e^{ t_i \psibar_i \psi_i}$.
We obtain
\begin{equation}
   \lambda (1-|A|) \tau_{A \smallsetminus i}
   \,+\, t_i \tau_{A \smallsetminus i}
   \,+\,\sum_{j \in A \smallsetminus i} \tau_{A \smallsetminus \{i,j\}}
   \,-\! \sum_{\begin{scarray}
                 j,k \in A \smallsetminus i \\
                 j \neq k
              \end{scarray}}
         \!
         \psibar_j \psi_k \tau_{A \smallsetminus \{i,j,k\}}
\end{equation}
(in the last term we must have $j,k \neq i$ by parity),
which equals $f_{A \smallsetminus i}^{(\lambda)}
                \,+\, (t_i-\lambda) \tau_{A \smallsetminus i}$ as claimed.
\qed

Applying Lemma~\ref{lemma.inti} repeatedly
for $i$ lying in an arbitrary set $B \subseteq V$, we obtain:

\begin{corollary}
  \label{cor.int}
Let $A,B \subseteq V$.  Then
\begin{equation}
   \int \mathcal{D}_{B,\bt}(\psi, \psibar) \, f_A^{(\lambda)}
  \;=\;
  \Biggl( \prod\limits_{i \in B \smallsetminus A} t_i \Biggr) \,
  \Biggl[ f_{A \smallsetminus B}^{(\lambda)} \,+\,
             \Bigl( \sum\limits_{i \in B \cap A} (t_i-\lambda) \Bigr)
             \tau_{A \smallsetminus B}
  \Biggr]
  \;.
 \label{eq.cor.int.1}
\end{equation}
In particular, for $B=A$ we have
\begin{equation}
   \int \mathcal{D}_{A,\bt}(\psi, \psibar) \, f_A^{(\lambda)}
  \;=\;
  \lambda \,+\, \sum\limits_{i \in A} (t_i-\lambda)
  \;.
 \label{eq.cor.int.2}
\end{equation}
\end{corollary}

\proof
The factors $t_i$ for $i \in B \setminus A$
follow trivially from the second line of Lemma~\ref{lemma.inti}(b).
For the rest, we proceed by induction on the cardinality of $B \cap A$.
If $|B \cap A| = 0$, the result is trivial.
So assume that the result holds for a given set $B$,
and consider $B' = B \cup \{j\}$ with $j \in A \setminus B$.
Using Lemma~\ref{lemma.inti}(a,b) we have
\begin{subeqnarray}
   & &
   \!\!\!\!
   \int \dx{\psi_j} \, \dx{\psibar_j} \, e^{ t_j \psibar_j \psi_j} \,
     \Biggl[ f_{A \smallsetminus B}^{(\lambda)} \,+\,
                \biggl( \sum\limits_{i \in B \cap A} (t_i-\lambda) \biggr)
                \tau_{A \smallsetminus B}
     \Biggr]
           \nonumber \\[1mm]
   & & \quad
   =\;
   f_{(A \smallsetminus B) \smallsetminus \{j\}}^{(\lambda)}
   \,+\, 
   (t_j-\lambda) \tau_{(A \smallsetminus B) \smallsetminus \{j\}}
   \,+\,
   \biggl( \sum\limits_{i \in B \cap A} (t_i-\lambda) \biggr)
             \tau_{(A \smallsetminus B) \smallsetminus \{j\}}
      \qquad \\[1mm]
   & & \quad
   =\;
   f_{A \smallsetminus B'}^{(\lambda)} \,+\,
                \biggl( \sum\limits_{i \in B' \cap A} (t_i-\lambda) \biggr)
                \tau_{A \smallsetminus B'}
   \;,
\end{subeqnarray}
as claimed.
\qed

Applying \reff{eq.cor.int.2} once for each factor $C_\alpha$,
we have:

\begin{corollary}
  \label{cor.int2}
Let $\{ C_\alpha \}$ be a partition of $V$.  Then
\begin{equation}
   \int \mathcal{D}_{V,\bt}(\psi, \psibar)
       \, \prod\limits_{\alpha} f_{C_\alpha}^{(\lambda_\alpha)}
  \;=\;
  \prod_{\alpha}   \Big( \lambda_\alpha +
                         \sum_{i \in C_{\alpha}} (t_i-\lambda_\alpha) \Big)
  \;.
 \label{eq.cor.int.partition}
\end{equation}
\end{corollary}

%

The partition function \reff{eq.Zgrass} can now be computed
immediately by combining Corollaries~\ref{cor.exp} and \ref{cor.int2}.
We obtain the main result of this section:

\begin{theorem}
   \label{thm.Zgrass}
Let $G=(V,E)$ be a hypergraph,
and let $\{ w_A \} _{A \in E}$ be hyperedge weights.
Then
\begin{eqnarray}
   & &
   \int \! \mathcal{D}(\psi, \psibar) \,
      \exp \Biggl[ \sum_i t_i \psibar_i \psi_i
                      \,+\,  \sum_{A \in E} w_A f_A^{(\lambda_A)} \Biggr]
       \nonumber \\
   & & 
   \qquad =\!\!
   \sum_{\begin{scarray}
              F\in \scrf(G)  \\
              F = (F_1,\ldots,F_{\ell})
         \end{scarray}}
   \!\!\!
     \Biggl( \prod\limits_{A \in F} w_A \! \Biggr)
   \prod_{\alpha=1}^{\ell}
        \Biggl(\, \sum_{i \in V(F_{\alpha})} t_i
              \,- \sum_{A \in E(F_{\alpha})} (|A|-1) \lambda_A \Biggr)
  \;, \qquad
 \label{eq.thm.Zgrass.gen}
\end{eqnarray}
where the sum runs over spanning hyperforests $F$ in $G$
with components $F_1,\ldots,F_{\ell}$,
and $V(F_\alpha)$ is the vertex set of the hypertree $F_\alpha$.
In particular, if $\lambda_A$ takes the same value for all $A$, we have
\begin{equation}
   \int \! \mathcal{D}(\psi, \psibar) \,
      \exp \Biggl[ \sum_i t_i \psibar_i \psi_i
                      \,+\,  \sum_{A \in E} w_A f_A^{(\lambda)} \Biggr]
   \,= \!\!\!\!
   \sum_{\begin{scarray}
              F\in \scrf(G)  \\
              F = (F_1,\ldots,F_{\ell})
         \end{scarray}}
   \!\!\!
     \Biggl( \prod\limits_{A \in F} w_A \! \Biggr)
   \prod_{\alpha=1}^{\ell}
        \Biggl( \lambda +\! \sum_{i \in V(F_{\alpha})} (t_i-\lambda) \Biggr)
  \;,
 \label{eq.thm.Zgrass}
\end{equation}
\end{theorem}

\proof
We apply \reff{eq.cor.int.partition},
where (according to Corollary~\ref{cor.exp})
$\lambda_\alpha$ is the weighted average \reff{eq.weighted_lambda}
taken over the hyperedges contained in the hypertree $F_\alpha$.
Then
\begin{subeqnarray}
   \lambda_\alpha +\! \sum_{i \in V(F_\alpha)} (t_i-\lambda_\alpha)
   & = &
   \sum_{i \in V(F_{\alpha})} t_i
              \,-\, \lambda_\alpha (|V(F_\alpha)| - 1)
       \\[2mm]
   & = &
   \sum_{i \in V(F_{\alpha})} t_i
              \,- \sum_{A \in E(F_{\alpha})} (|A|-1) \lambda_A
   \;.
\end{subeqnarray}
\qed

If we specialize \reff{eq.thm.Zgrass} to $t_i = \lambda$ for all vertices $i$,
we obtain:

\begin{corollary}
   \label{cor1.Zgrass}
Let $G=(V,E)$ be a hypergraph,
and let $\{ w_A \} _{A \in E}$ be hyperedge weights.
Then
\begin{subeqnarray}
   \!\!
   \int \! \mathcal{D}(\psi, \psibar) \,
      \exp \Biggl[ \lambda \sum_i \psibar_i \psi_i
                      \,+\,  \sum_{A \in E} w_A f_A^{(\lambda)} \Biggr]
   & = & \!\!
   \sum_{F\in \scrf(G)}
   \!
     \Biggl( \prod\limits_{A \in F} w_A \! \Biggr)
   \; \lambda^{k(F)}
         \\[2mm]
   & = &
   \lambda^{|V|}
   \!\!
   \sum_{F\in \scrf(G)}
   \!
     \Biggl( \prod\limits_{A \in F} \frac{w_A}{\lambda^{|A|-1}} \! \Biggr)
   \;, \qquad\quad
 \label{eq.cor1.Zgrass}
\end{subeqnarray}
where the sum runs over spanning hyperforests $F$ in $G$,
and $k(F)$ is the number of connected components of $F$.
\end{corollary}

\noindent
This is the generating function of {\em unrooted}\/ spanning hyperforests,
with a weight $w_A$ for each hyperedge $A$
and a weight $\lambda$ for each connected component.
Note that the second equality in \reff{eq.cor1.Zgrass}
uses Proposition~\ref{prop.eulerineq.hypergraphs}.

If, on the other hand, we specialize \reff{eq.thm.Zgrass} to $\lambda = 0$,
we obtain:

\begin{corollary}
   \label{cor2.Zgrass}
Let $G=(V,E)$ be a hypergraph,
and let $\{ w_A \} _{A \in E}$ be hyperedge weights.
Then
\begin{equation}
   \int \! \mathcal{D}(\psi, \psibar) \,
      \exp \Biggl[ \sum_i t_i \psibar_i \psi_i
                      \,+\,  \sum_{A \in E} w_A f_A^{(0)} \Biggr]
   \;= \!\!
   \sum_{\begin{scarray}
              F\in \scrf(G)  \\
              F = (F_1,\ldots,F_{\ell})
         \end{scarray}}
   \!\!\!
     \Biggl( \prod\limits_{A \in F} w_A \! \Biggr)
   \prod_{\alpha=1}^{\ell}
        \biggl( \sum_{i \in V(F_{\alpha})} t_i \biggr)
  \;,
 \label{eq.cor2.Zgrass}
\end{equation}
where the sum runs over spanning hyperforests $F$ in $G$
with components $F_1,\ldots,F_{\ell}$,
and $V(F_\alpha)$ is the vertex set of the hypertree $F_\alpha$.
\end{corollary}

\noindent
This is the generating function of {\em rooted}\/ spanning hyperforests,
with a weight $w_A$ for each hyperedge $A$
and a weight $t_i$ for each root $i$.

Finally, returning to the case in which $t_i = \lambda$ for all $i$,
we can obtain a formula more general than \reff{eq.cor1.Zgrass}
in which the left-hand side contains an additional factor
$f_{\scrc}^{(\lambda)} = \prod\limits_\gamma f_{C_\gamma}^{(\lambda)}$,
where $\scrc = \{ C_\gamma \}$ is an arbitrary family 
of disjoint nonempty subsets of $V$.
Indeed, it suffices to differentiate \reff{eq.cor1.Zgrass}
with respect to all the weights $w_{C_\gamma}$.\footnote{
   If the sets $C_\gamma$ do not happen to belong to the hyperedge set $E$,
   it suffices to adjoin them to $E$ and give them weight $w_{C_\gamma} = 0$.
   Indeed, there is no loss of generality in assuming that $G$ is
   the complete hypergraph on the vertex set $V$,
   i.e.\ that {\em every}\/ subset of $V$ of cardinality $\ge 2$
   is a hyperedge.
}
We obtain:

\begin{corollary}
   \label{cor1.Zgrass.bis}
Let $G=(V,E)$ be a hypergraph,
let $\{ w_A \} _{A \in E}$ be hyperedge weights,
and let $\scrc = \{ C_\gamma \}$ be a family of
disjoint nonempty subsets of $V$.
Then
\begin{eqnarray}
   & &
   \int \! \mathcal{D}(\psi, \psibar) \,
   \Biggl( \prod\limits_\gamma f_{C_\gamma}^{(\lambda)}  \Biggr) \,
      \exp \Biggl[ \lambda \sum_i \psibar_i \psi_i
                      \,+\,  \sum_{A \in E} w_A f_A^{(\lambda)} \Biggr]
          \qquad\qquad \nonumber \\[2mm]
   & &
   \qquad\qquad =\;
   \sum_{F\in \scrf(G;\scrc)}
   \!
     \Biggl( \prod\limits_{A \in F} w_A \! \Biggr)
   \; \lambda^{k(F) - \sum\limits_{\gamma} (|C_\gamma|-1)}
   \;,
 \label{eq.cor1.Zgrass.bis}
\end{eqnarray}
where $\scrf(G;\scrc)$ denotes the set of spanning hyperforests in $G$
that do not contain any of the $\{ C_\gamma \}$ as hyperedges
and that remain hyperforests (i.e., acyclic)
when the hyperedges $\{ C_\gamma \}$ are adjoined.
\end{corollary}

\noindent
Indeed, to deduce Corollary~\ref{cor1.Zgrass.bis}
from Corollary~\ref{cor1.Zgrass} by differentiation,
it suffices to observe that, by Proposition~\ref{prop.eulerineq.hypergraphs},
the number of connected components in the hyperforest obtained from $F$
by adjoining the hyperedges $\{ C_\gamma \}$
is precisely $k(F) - \sum\limits_{\gamma} (|C_\gamma|-1)$.

For instance, if $\prod\limits_\gamma f_{C_\gamma}^{(\lambda)}$
consists of a single factor $f_C$,
then $\scrf(G;\{C\})$ consists of the spanning hyperforests
in which all the vertices of the set $C$ belong to different components.
Similarly, if $\prod\limits_\gamma f_{C_\gamma}^{(\lambda)}$
consists of two factors $f_{C_1} f_{C_2}$ with $C_1 \cap C_2 = \emptyset$,
then $\scrf(G;\{C_1,C_2\})$ consists of the spanning hyperforests
in which each component contains at most one vertex from $C_1$
and at most one vertex from $C_2$.
The conditions get somewhat more complicated when
there are three or more sets $C_\gamma$.

It is possible to obtain an analogous extension of
Theorem~\ref{thm.Zgrass} by the same method,
but the weights get somewhat complicated,
precisely because we lose the opportunity of using
Proposition~\ref{prop.eulerineq.hypergraphs} in a simple way.

Equations~\reff{eq.thm.Zgrass.gen}--\reff{eq.cor2.Zgrass}
are the hypergraph generalization of
\reff{eq.fourfermion.1}--\reff{eq.fourfermion.4}, respectively.
To see this, let $G=(V,E)$ be an ordinary graph,
so that each edge $e \in E$ is simply an unordered pair $\{i,j\}$
of distinct vertices $i,j \in V$,
to which there is associated an edge weight $w_{ij} = w_{ji}$.
Then by definition~\reff{eq.deff_A} we have
\begin{equation}
f_{\{i,j\}}^{(\lambda)}(\psi, \psibar) 
   \;=\;
  - \lambda \,\psibar_i\psi_i \psibar_j\psi_j +
  \psibar_i \psi_i + \psibar_j \psi_j - \psibar_i \psi_j - \psibar_j \psi_i
   \;,
\end{equation}
so that if we take $\lambda_{ij} = -u_{ij}$ we have
\begin{equation}
\sum_{\{i,j\}\in E} w_ {ij} f_ {\{i,j\}}^{(-u_{ij})} =
   \sum_{i,j\in V} \psibar_i L_{ij} \psi_j  \,+\,
   \sum_{\{i,j\}\in E} u_{ij} w_{ij} \psibar_i\psi_i \psibar_j\psi_j 
\end{equation}
where the {\em (weighted) Laplacian matrix}\/ for the graph $G$ is defined as
\begin{equation}
L_{ij} \;=\; \begin{cases}
                - w_{ij} & \hbox{if } i \neq j \\
                \sum\limits_{k \neq i} w_{ik} & \hbox{if }  i=j
             \end{cases}
 \label{eq.Laplacian.graph}
\end{equation}
Then \reff{eq.thm.Zgrass.gen}--\reff{eq.cor2.Zgrass}
become precisely \reff{eq.fourfermion.1}--\reff{eq.fourfermion.4},
respectively.


More generally, consider the case in which $G=(V,E)$
is a $k$-uniform hypergraph
(an ordinary graph corresponds to the case $k=2$).
Let $w_{i_1,\ldots,i_k}$ (assumed completely symmetric in the
indices $i_1,\ldots,i_k$) be the weight associated to the hyperedge
$\{i_1,\ldots,i_k\}$ when $i_1,\ldots,i_k$ are all distinct,
and let $w_{i_1,\ldots,i_k}=0$ when at least two indices are equal.
Define the {\em (weighted) Laplacian tensor}\/
(a rank-$k$ symmetric tensor) by
\begin{equation}
L_{i_1,\ldots,i_k}   \;=\;
  \begin{cases}
    - w_{i_1,\ldots,i_k} & \hbox{if $i_1,\ldots,i_k$ are all different}
                 \\[2mm]
    \frac{1}{k-1} \sum\limits_{i'_s} w_{i_1,\ldots,i'_s,\ldots,i_k}
       & \hbox{if $i_r=i_s$ ($r\neq s$) and the others are all different}
                 \\[2mm]
    0  & \hbox{otherwise}
  \end{cases}
 \label{eq.Laplacian.k-uniform}
\end{equation}
Then we have
\begin{equation}
  \sum_{A \in E} w_A f_A^{(\lambda)}
   \;=\;
   \sum_{i_1,\ldots,i_k\in V}  \frac{L_{i_1,\ldots,i_k}}{(k-2)!}
          \left[\psibar_{i_1} \psi_{i_2} \psibar_{i_3} \psi_{i_3} \cdots 
                   \psibar_{i_k}\psi_{i_k} 
                \,+\,
                \frac{\lambda}{k}
                \psibar_{i_1}\psi_{i_1}\cdots \psibar_{i_k}\psi_{i_k}
          \right]
  \;,
 \label{eq.k-uniform}
\end{equation}
so that the ``action'' is given by \reff{eq.k-uniform}
plus the ``mass term'' $\lambda \sum_i \psibar_i \psi_i$.
Combining Corollary~\ref{cor1.Zgrass} with \reff{eq.k-uniform},
we obtain a formula for the generating function of spanning hyperforests
in a $k$-uniform hypergraph:
\begin{eqnarray}
   & & \!\!\!\!
   \sum_{F\in \scrf(G)}
      \!
        \Biggl( \prod\limits_{A \in F} w_A \! \Biggr)
      \; \lambda^{k(F)}
   \;\,=\;\,
   \int \! \mathcal{D}(\psi, \psibar) \,\times
       \nonumber  \\
   & & \quad
      \exp \Biggl\{ \lambda \sum_i \psibar_i \psi_i
                      \,+\!\!
   \sum_{i_1,\ldots,i_k\in V}  \frac{L_{i_1,\ldots,i_k}}{(k-2)!}
          \biggl[\psibar_{i_1} \psi_{i_2} \psibar_{i_3} \psi_{i_3} \cdots
                   \psibar_{i_k}\psi_{i_k}
                \,+\,
                \frac{\lambda}{k}
                \psibar_{i_1}\psi_{i_1}\cdots \psibar_{i_k}\psi_{i_k}
          \biggr]
          \Biggr\}
   \;.
       \nonumber  \\
 \label{cor1.Zgrass.k-uniform}
\end{eqnarray}

Let us remark that while the Laplacian matrix \reff{eq.Laplacian.graph}
for an ordinary graph has vanishing row and column sums
(i.e., $\sum_j L_{ij} = 0$),
the Laplacian tensor \reff{eq.Laplacian.k-uniform} for a hypergraph
satisfies $\sum_{i_k} L_{i_1,\ldots,i_k} = 0$
when $i_1,\ldots,i_{k-1}$ are all distinct,
but not in general otherwise.

For an application to counting spanning hyperforests
in the complete $k$-uniform hypergraph,
see \cite{Bedini}.


\bigskip

{\bf Remark.}
One can also generalize \reff{eq.Zgrass} to allow products
$f_\scrc^{(\lambda)} = \prod\limits_\alpha f_{C_\alpha}^{(\lambda)}$
in the exponential (i.e., in the action)
in place of the single operators $f_A^{(\lambda)}$,
with corresponding coefficients $w_\scrc$.
These generalized integrals likewise lead to
polynomials in the variables $\{w_\scrc\}$
such that the union of the families $\scrc_j$ arising
in any given monomial is the set of hyperedges of a hyperforest.
However, the simultaneous presence of certain sets of hyperedges
in the hyperforest now gets extra weights.
We hope to discuss these extensions elsewhere.
This generalized model is conceptually important because,
when $t_i = \lambda$ for all $i$,
it corresponds to the {\em most general}\/ $\OSP(1|2)$-invariant action
(see Section \ref{sec:osp}).

\section{Extension to correlation functions}
\label{sec:corrfn}

In the preceding section we saw how the partition function \reff{eq.Zgrass}
of a particular class of fermionic theories
can be given a combinatorial interpretation
as an expansion over spanning hyperforests in a hypergraph.
In this section we will extend this result to give a
combinatorial interpretation for a class of Grassmann integrals
that correspond to (unnormalized) correlation functions
in this same fermionic theory;
we will obtain a sum over partially rooted spanning hyperforests
satisfying particular connection conditions.

Given ordered $k$-tuples of vertices
$I=(i_1,i_2,\dots, i_k) \in V^k$ and
$J=(j_1,j_2,\dots, j_k) \in V^k$,
let us define the operator
\begin{equation}
  \label{eq.operO}
\mathcal{O}_{I,J}
  \; := \;
\psibar_{i_1} \psi_{j_1} \cdots 
\psibar_{i_k} \psi_{j_k}
   \;,
\end{equation}
which is an even element of the Grassmann algebra.
Of course, the $i_1,i_2,\dots, i_k$ must be all distinct,
as must the $j_1,j_2,\dots, j_k$, or else we will have $\mathcal{O}_{I,J} = 0$.
We shall therefore assume henceforth that $I,J \in V_{\neq}^k$,
where $V_{\neq}^k$ is the set of ordered $k$-tuples of
{\em distinct}\/ vertices in $V$.
Note, however, that there can be overlaps between the sets
$\{i_1,i_2,\dots, i_k\}$ and $\{j_1,j_2,\dots, j_k\}$.
Note finally that $\mathcal{O}_{I,J}$ is antisymmetric under
permutations of the sequences $I$ and $J$, in the sense that
\begin{equation}
   \mathcal{O}_{I \circ \sigma, J \circ \tau}
   \;=\;
   \sgn(\sigma) \, \sgn(\tau) \, \mathcal{O}_{I,J}
\end{equation}
for any permutations $\sigma,\tau$ of $\{1,\ldots,k\}$.

Our goal in this section is to provide a combinatorial interpretation,
in terms of partially rooted spanning hyperforests
satisfying suitable connection conditions, for the general
Grassmann integral (``unnormalized correlation function'')
\begin{subeqnarray}
   [\mathcal{O}_{I,J}]  \;=\;  Z \langle \mathcal{O}_{I,J} \rangle
   & = &
   \int \mathcal{D}(\psi, \psibar) \,
      \mathcal{O}_{I,J} \,
      \exp \Biggl[ \sum_i t_i \psibar_i \psi_i
                      \,+\,  \sum_{A \in E} w_A f_A^{(\lambda)} \Biggr]
         \qquad \\[2mm]
   & = &
   \int \mathcal{D}_{V,\bt}(\psi, \psibar) \,
      \mathcal{O}_{I,J} \,
      \exp \Biggl[ \sum_{A \in E} w_A f_A^{(\lambda)} \Biggr]
   \;.
 \label{eq.corrfn.grass}
\end{subeqnarray}
The principal tool is the following generalization of \reff{eq.cor.int.2}:

\begin{lemma}
  \label{lemma.corrfn1}
Let $A \subseteq V$,
and let $I=(i_1,i_2,\dots, i_k) \in A_{\neq}^k$ and
$J=(j_1,j_2,\dots, j_k) \in A_{\neq}^k$.  Then
\begin{equation}
   \int \mathcal{D}_{A,\bt}(\psi, \psibar)
       \, \mathcal{O}_{I,J} \, f_A^{(\lambda)}
  \;=\;
   \begin{cases}
       \lambda \,+\, \sum\limits_{i \in A} (t_i-\lambda)
                                            & \textrm{if } k=0  \\
       1                                    & \textrm{if } k=1  \\[2mm]
       0                                    & \textrm{if } k \ge 2
   \end{cases}
\end{equation}
\end{lemma}

\proof
The case $k=0$ is just \reff{eq.cor.int.2}.
To handle $k=1$, recall that
\begin{equation}
f_A^{(\lambda)}   \; = \;
\lambda (1-|A|) \tau_A \,+\, \sum_{\ell \in A} \tau_{A \smallsetminus l}
  \,-\! \sum_{\begin{scarray}
                 \ell, m \in A \\
                 \ell \neq m
              \end{scarray}}
         \!
         \psibar_{\ell} \psi_m \tau_{A \smallsetminus \{l,m\}}
\ef.
\end{equation}
Now multiply $f_A^{(\lambda)}$ by $\psibar_i \psi_j$ with $i,j \in A$,
and integrate with respect to $\mathcal{D}_{A,\bt}(\psi, \psibar)$.
If $i=j$, then the only nonzero contribution comes from the term $\ell=i$
in the single sum, and $\psibar_i \psi_i \tau_{A \smallsetminus i} = \tau_A$,
so the integral is 1.
If $i \neq j$, then the only nonzero contribution comes from the term
$\ell=j$, $m=i$ in the double sum,
and $(\psibar_i \psi_j)(-\psibar_j \psi_i) \tau_{A \smallsetminus \{i,j\}}
     = \tau_A$,
so the integral is again 1.

Finally, if $|I| = |J| = k \ge 2$, then every monomial in
$\mathcal{O}_{I,J}  f_A^{(\lambda)}$
has degree $\ge 2|A| - 2 + 2k > 2|A|$,
so $\mathcal{O}_{I,J}  f_A^{(\lambda)} = 0$.
\qed

Of course, it goes without saying that if $m(\psi,\psibar)$
is a monomial of degree $k$ in the variables $\psi_i$ ($i \in A$)
and degree $k'$ in the variables $\psibar_i$ ($i \in A$),
and $k$ is {\em not}\/ equal to $k'$, then
$\int \! \mathcal{D}_{A,\bt}(\psi, \psibar) \, m(\psi,\psibar) \,
  f_A^{(\lambda)} = 0$.

Now go back to the general case
$I=(i_1,i_2,\dots, i_k) \in V_{\neq}^k$ and
$J=(j_1,j_2,\dots, j_k) \in V_{\neq}^k$,
let $\scrc = \{C_{\alpha} \}_{\alpha=1}^m$ be a partition of $V$,
and consider the integral
\begin{equation}
 \label{eq.IPIJdef}
   \mathcal{I}(I,J; \scrc)
   \;  := \;
   \int \mathcal{D}_{V,\bt}(\psi, \psibar) \,
      \mathcal{O}_{I,J} \, \prod_{\alpha=1}^m f_{C_{\alpha}}^{(\lambda)}
\ef.
\end{equation}
The integral factorizes over the sets $C_{\alpha}$ of the partition,
and it vanishes unless $|I \cap C_\alpha| = |J \cap C_\alpha|$
for all $\alpha$;
here $I \cap C_\alpha$ denotes the subsequence of $I$ consisting of those
elements that lie in $C_\alpha$, kept in their original order,
and $|I \cap C_\alpha|$ denotes the length of that subsequence
(and likewise for $J \cap C_\alpha$).
So let us decompose the operator $\mathcal{O}_{I,J}$ as
\begin{equation}
\mathcal{O}_{I,J}=
\sigma(I,J; \scrc)
\prod_{\alpha=1}^m
\mathcal{O}_{I \cap C_{\alpha}, J \cap C_{\alpha}}
\ef,
\end{equation}
where $\sigma(I,J; \scrc) \in \{\pm 1\}$ is a sign coming from the
reordering of the fields in the product.
Applying Lemma~\ref{lemma.corrfn1} once for each factor $C_\alpha$,
we see that the integral \reff{eq.IPIJdef} is nonvanishing
only if $|I \cap C_\alpha| = |J \cap C_\alpha| \le 1$ for all $\alpha$:
that is, each set $C_\alpha$ must contain either one element from $I$
and one element from $J$ (possibly the same element)
or else no element from $I$ or $J$.
Let us call the partition $\scrc$ {\em properly matched}\/ for $(I,J)$
when this is the case.
(Note that this requires in particular that $m \ge k$.)
Note also that for properly matched partitions $\scrc$
we can express the combinatorial sign $\sigma(I,J; \scrc)$ in a simpler way:
it is the sign of the unique permutation $\pi$ of $\{1,\ldots,k\}$
such that $i_r$ and $j_{\pi(r)}$
lie in the same set $C_\alpha$ for each $r$ ($1 \le r \le k$).
(Note in particular that when
$\{i_1,i_2,\dots, i_k\} \cap \{j_1,j_2,\dots, j_k\} \equiv S \neq \emptyset$,
the pairing $\pi$ has to match the repeated elements
[i.e., $i_r = j_{\pi(r)}$ whenever $i_r \in S$],
since a vertex cannot belong simultaneously to two distinct blocks
$C_\alpha$ and $C_\beta$.)
We then deduce immediately from Lemma~\ref{lemma.corrfn1}
the following generalization of Corollary~\ref{cor.int2}:

\begin{corollary}
  \label{cor.int2.corrfn}
Let $I,J \in V_{\neq}^k$ and let $\scrc = \{ C_\alpha \}$ be a partition of $V$.
Then
\begin{eqnarray}
 \label{eq.cor.int.partition.corrfn}
 & & \!\!\!\!
   \int \mathcal{D}_{V,\bt}(\psi, \psibar) \, \mathcal{O}_{I,J}
       \, \prod\limits_{\alpha} f_{C_\alpha}^{(\lambda)}
       \nonumber \\
 & & =\; 
   \begin{cases}
      \sgn(\pi) \!
      \prod\limits_{\alpha\colon |I \cap C_\alpha| = 0}  \!
      \Big( \lambda + \sum\limits_{i \in C_{\alpha}} (t_i-\lambda) \Big)
         & \textrm{if $\scrc$ is properly matched for $(I,J)$} \\
      0  & \textrm{otherwise}
   \end{cases}
\end{eqnarray}
where $\pi$ is the permutation of $\{1,\ldots,k\}$
such that $i_r$ and $j_{\pi(r)}$
lie in the same set $C_\alpha$ for each $r$.
\end{corollary}


We can now compute the integral \reff{eq.corrfn.grass}
by combining Corollaries~\ref{cor.exp} and \ref{cor.int2.corrfn}.
If $G=(V,E)$ is a hypergraph and $G'$ is a spanning subhypergraph of $G$,
let us say that $G'$ is {\em properly matched}\/ for $(I,J)$
[we denote this by $G' \sim (I,J)$]
in case the partition of $V$ induced by the decomposition of $G'$
into connected components is properly matched for $(I,J)$.
We then obtain the main result of this section:

\begin{theorem}
   \label{thm.corrfn.grass}
Let $G=(V,E)$ be a hypergraph,
let $\{ w_A \} _{A \in E}$ be hyperedge weights,
and let $I,J \in V_{\neq}^k$.
Then
\begin{eqnarray}
   & & \!\!\!
   \int \! \mathcal{D}(\psi, \psibar) \, \mathcal{O}_{I,J} \,
      \exp \Biggl[ \sum_i t_i \psibar_i \psi_i
                      \,+\,  \sum_{A \in E} w_A f_A^{(\lambda)} \Biggr]
       \nonumber \\
   & & \;
   =\!\!\!
   \sum_{\begin{scarray}
              F\in \scrf(G)  \\
              F \sim (I,J) \\
              F = (F_1,\ldots,F_{\ell}) \\
         \end{scarray}}
   \!\!\!\!\!
   \sgn(\pi_{I,J;F}) \,
     \Biggl( \prod\limits_{A \in F} w_A \! \Biggr)
   \!
   \prod_{\alpha \colon |I \cap V(F_{\alpha})| = 0}
        \biggl( \lambda + \sum_{i \in V(F_{\alpha})} (t_i-\lambda) \biggr)
  \;,
  \qquad
 \label{eq.thm.corrfn.grass}
\end{eqnarray}
where the sum runs over spanning hyperforests $F$ in $G$,
with components $F_1,\ldots,F_{\ell}$,
that are properly matched for $(I,J)$,
and $V(F_\alpha)$ is the vertex set of the hypertree $F_\alpha$;
here $\pi_{I,J;F}$ is the permutation of $\{1,\ldots,k\}$
such that $i_r$ and $j_{\pi(r)}$
lie in the same component $F_\alpha$ for each $r$.
\end{theorem}

If we specialize \reff{eq.thm.corrfn.grass}
to $t_i = \lambda$ for all vertices $i$, we obtain:

\begin{corollary}
   \label{cor1.corrfn.grass}
Let $G=(V,E)$ be a hypergraph,
let $\{ w_A \} _{A \in E}$ be hyperedge weights.
and let $I,J \in V_{\neq}^k$.
Then
\begin{subeqnarray}
   & &
   \int \! \mathcal{D}(\psi, \psibar) \, \mathcal{O}_{I,J} \,
      \exp \Biggl[ \lambda \sum_i \psibar_i \psi_i
                      \,+\,  \sum_{A \in E} w_A f_A^{(\lambda)} \Biggr]
       \nonumber \\
   & & \qquad\qquad
   =\!
   \sum_{\begin{scarray}
              F\in \scrf(G)  \\
              F \sim (I,J)
         \end{scarray}}
   \!\!
   \sgn(\pi_{I,J;F}) \,
     \Biggl( \prod\limits_{A \in F} w_A \! \Biggr)
   \; \lambda^{k(F)-k}
         \\[2mm]
   & & \qquad\qquad
   =\;
   \lambda^{|V|-k}
   \!\!
   \sum_{\begin{scarray}
              F\in \scrf(G)  \\
              F \sim (I,J)
         \end{scarray}}
   \!\!
   \sgn(\pi_{I,J;F}) \,
     \Biggl( \prod\limits_{A \in F} \frac{w_A}{\lambda^{|A|-1}} \! \Biggr)
   \;, \qquad\quad
 \label{eq.cor1.corrfn.grass}
\end{subeqnarray}
where the sum runs over spanning hyperforests $F$ in $G$
that are properly matched for $(I,J)$,
and $k(F)$ is the number of connected components of $F$;
here $\pi_{I,J;F}$ is the permutation of $\{1,\ldots,k\}$
such that $i_r$ and $j_{\pi(r)}$
lie in the same component of $F$ for each $r$.
\end{corollary}

\noindent
This is the generating function of spanning hyperforests
that are rooted at the vertices in $I,J$ and are otherwise unrooted,
with a weight $w_A$ for each hyperedge $A$
and a weight $\lambda$ for each unrooted connected component.

If, on the other hand, we specialize \reff{eq.thm.corrfn.grass}
to $\lambda = 0$, we obtain:

\begin{corollary}
   \label{cor2.corrfn.grass}
Let $G=(V,E)$ be a hypergraph,
let $\{ w_A \} _{A \in E}$ be hyperedge weights,
and let $I,J \in V_{\neq}^k$.
Then
\begin{eqnarray}
   & &
   \int \! \mathcal{D}(\psi, \psibar) \, \mathcal{O}_{I,J} \,
      \exp \Biggl[ \sum_i t_i \psibar_i \psi_i
                      \,+\,  \sum_{A \in E} w_A f_A^{(0)} \Biggr]
           \nonumber \\
   & & \qquad
   =\!\!
   \sum_{\begin{scarray}
              F\in \scrf(G)  \\
              F \sim (I,J) \\
              F = (F_1,\ldots,F_{\ell})
         \end{scarray}}
   \!\!
   \sgn(\pi_{I,J;F}) \,
     \Biggl( \prod\limits_{A \in F} w_A \! \Biggr)
   \!
   \prod_{\alpha \colon |I \cap V(F_{\alpha})| = 0}
        \biggl( \sum_{i \in V(F_{\alpha})} t_i \biggr)
  \;,
  \qquad
 \label{eq.cor2.corrfn.grass}
\end{eqnarray}
where the sum runs over spanning hyperforests $F$ in $G$,
with components $F_1,\ldots,F_{\ell}$,
that are properly matched for $(I,J)$,
and $V(F_\alpha)$ is the vertex set of the hypertree $F_\alpha$;
here $\pi_{I,J;F}$ is the permutation of $\{1,\ldots,k\}$
such that $i_r$ and $j_{\pi(r)}$
lie in the same component $F_\alpha$ for each $r$.
\end{corollary}

\noindent
This is the generating function of rooted spanning hyperforests,
with a weight $w_A$ for each hyperedge $A$
and a weight $t_i$ for each root $i$
{\em other than}\/ those in the sets $I,J$.

Let us conclude by making some remarks about the
{\em normalized}\/ correlation function $\langle \mathcal{O}_{I,J} \rangle$
obtained by dividing \reff{eq.corrfn.grass} by \reff{eq.Zgrass}.
For simplicity, let us consider only the two-point function
$\langle \psibar_i \psi_j \rangle$.
We have
\begin{equation}
   \langle \psibar_i \psi_j \rangle
   \;=\;
   \left\langle
      \gamma_{ij} \,
      \Big( \lambda + \sum\limits_{k \in \Gamma(i)} (t_k-\lambda) \Big)^{-1}
   \right\rangle
   \;,
 \label{eq.2pointfn}
\end{equation}
where the expectation value on the right-hand side
is taken with respect to the ``probability distribution''\footnote{
   We write ``probability distribution'' in quotation marks
   because the ``probabilities'' will in general be complex.
   They will be true probabilities (i.e., real numbers between 0 and 1)
   if the hyperedge weights $w_A$ are nonnegative real numbers.
}
on spanning hyperforests of $G$
in which the hyperforest $F = (F_1,\ldots,F_{\ell})$ gets weight
\begin{equation}
    Z^{-1} \,
    \Biggl( \prod\limits_{A \in F} w_A \! \Biggr)
   \prod_{\alpha=1}^{\ell}
        \Big( \lambda + \sum_{k \in V(F_{\alpha})} (t_k-\lambda) \Big)
    \;,
\end{equation}
$\gamma_{ij}$ denotes the indicator function
\begin{equation}
   \gamma_{ij}
   \;=\;
   \begin{cases}
       1  & \textrm{if $i$ and $j$ belong to the same component of $F$} \\
       0  & \textrm{if not}
   \end{cases}
\end{equation}
and $\Gamma(i)$ denotes the vertex set of the component of $F$ containing $i$.
The factor
$\Big( \lambda + \sum\limits_{k \in \Gamma(i)} (t_k-\lambda) \Big)^{-1}$
in \reff{eq.2pointfn} arises from the fact that in \reff{eq.thm.Zgrass}
each component gets a weight
$\lambda + \sum_{k \in \Gamma(i)} (t_k-\lambda)$,
while in \reff{eq.thm.corrfn.grass} only those components {\em other than}\/
the one containing $i$ and $j$ get such a weight.
So in general the correlation function $\langle \psibar_i \psi_j \rangle$
is {\em not}\/ simply equal to (or proportional to)
the connection probability $\langle \gamma_{ij} \rangle$.
However, in the special case of
Corollaries~\ref{cor1.Zgrass} and \ref{cor1.corrfn.grass}
--- namely, all $t_i = \lambda$, so that we get unrooted spanning
hyperforests with a ``flat'' weight $\lambda$ for each component ---
then we have the simple identity
\begin{equation}
   \langle \psibar_i \psi_j \rangle
   \;=\;
   \lambda^{-1}  \langle \gamma_{ij} \rangle
   \;.
 \label{eq.2pointfn.special}
\end{equation}
Combinatorial identities generalizing \reff{eq.2pointfn.special},
and their relation to
the Ward identities arising from the $\OSP(1|2)$ supersymmetry,
will be discussed elsewhere \cite{us_Ward}.

\section{The role of $\OSP(1|2)$ symmetry}
\label{sec:osp}

In~\cite{us_prl} we have shown how the fermionic 
theory \reff{eq.fourfermion.3}
emerges naturally from the expansion of a theory with
bosons and fermions taking values in the unit supersphere in $\R^{1|2}$,
when the action is quadratic and invariant under rotations in $\OSP(1|2)$.
Here we would like to discuss this fact in greater detail,
and extend it to the hypergraph fermionic model \reff{eq.cor1.Zgrass}.

We begin by introducing, at each vertex $i \in V$,
a superfield $\nv_i := (\sigma_i,\psi_i,\psibar_i)$
consisting of a bosonic (i.e., real) variable $\sigma_i$ 
and a pair of Grassmann variables $\psi_i, \psibar_i$.
We equip the ``superspace'' $\R^{1|2}$ with the scalar product
\begin{equation}
   \nv_i\cdot\nv_j
   \; := \;
  \sigma_i \sigma_j \,+\, \lambda (\psibar_i \psi_j - \psi_i \psibar_j)
   \;,
 \label{eq.scalarprod}
\end{equation}
where $\lambda \neq 0$ is an arbitrary real parameter.

The infinitesimal rotations in $\R^{1|2}$
that leave invariant the scalar product \reff{eq.scalarprod}
form the Lie superalgebra $\osp(1|2)$
\cite{Rittenberg_78,Scheunert_79,Tolstoy}.
This algebra is generated by two types of transformations:
Firstly, we have the elements of the $\ssp(2)$ subalgebra,
which act on the field as $\nv'_i = \nv_i + \delta \nv_i$ with
\begin{subeqnarray}
 \delta \sigma_i & = &
    0  \\
 \delta \psi_i & = &
   -\, \alpha\, \psi_i + \gamma\, \psibar_i     \\
 \delta \psibar_i & = &
  +\, \alpha\, \psibar_i + \beta\, \psi_i
 \label{def.sp2}
\end{subeqnarray}
where $\alpha,\beta,\gamma$ are bosonic (Grassmann-even) global parameters;
it is easily checked that these transformations leave
\reff{eq.scalarprod} invariant.
Secondly, we have the transformations parametrized by
fermionic (Grassmann-odd) global parameters $\epsilon,\epsilonbar$:
\begin{subeqnarray}
 \delta \sigma_i & = &
    - \lambda^{1/2} (\epsilonbar \psi_i  +  \psibar_i \epsilon)  \\
 \delta \psi_i & = &
    \lambda^{-1/2}\,\epsilon \,\sigma_i     \\
 \delta \psibar_i & = &
    \lambda^{-1/2}\,\epsilonbar \,\sigma_i
 \label{def.supersym}
\end{subeqnarray}
(Here an overall  factor $\lambda^{-1/2}$ has been extracted
 from the fermionic parameters for future convenience.)
To check that these transformations leave \reff{eq.scalarprod} invariant,
we compute
\begin{subeqnarray}
   \delta( \nv_i\cdot\nv_j )
   & = &
   (\delta \sigma_i) \sigma_j + \sigma_i (\delta \sigma_j)
   + \lambda \big[ (\delta \psibar_i) \psi_j + \psibar_i (\delta \psi_j)
                  -(\delta \psi_i) \psibar_j - \psi_i (\delta \psibar_j) \big]
      \nonumber \\ \\[-1mm]
   & = &
   -\lambda^{1/2} (\epsilonbar \psi_i +  \psibar_i \epsilon) \sigma_j
   -\lambda^{1/2} (\epsilonbar \psi_j +  \psibar_j \epsilon) \sigma_i
           \nonumber \\
   &  & \qquad
   +\, \lambda^{1/2} \big[ \epsilonbar \psi_j \sigma_i
                          +  \psibar_i \epsilon \sigma_j
              -\epsilon \psibar_j \sigma_i - \psi_i \epsilonbar \sigma_j \big]
      \\[2mm]
   & = &
   0  \;.
\end{subeqnarray}

In terms of the differential operators
$\partial_i = \partial/\partial \psi_i$
and $\bar{\partial}_i = \partial/\partial \psibar_i$,
the transformations \reff{def.sp2} can be represented by the generators
\begin{subeqnarray}
 X_0  & = & \sum_{i\in V} (\psibar_i \bar{\partial}_i - \psi_i \partial_i)
   \\[1mm]
 X_+  & = & \sum_{i\in V} \psibar_i \partial_i
   \\[1mm]
 X_-  & = & \sum_{i\in V} \psi_i \bar{\partial}_i
 \label{eq:repsp2}
\end{subeqnarray}
corresponding to the parameters $\alpha,\beta,\gamma$, respectively,
while the transformations \reff{def.supersym}
can be represented by the generators
\begin{subeqnarray}
  Q_+  & = &   \lambda^{-1/2} \sum_{i\in V}  \sigma_i \partial_i 
               \,+\,
       \lambda^{1/2} \sum_{i\in V} \psibar_i \frac{\partial}{\partial \sigma_i}
   \\[1mm]
  Q_-  & = &   \lambda^{-1/2} \sum_{i\in V}  \sigma_i \bar{\partial}_i 
               \,-\,
       \lambda^{1/2} \sum_{i\in V} \psi_i \frac{\partial}{\partial \sigma_i}
 \label{eq:def_Q}
\end{subeqnarray}
corresponding to the parameters $\epsilon,\epsilonbar$, respectively.
(With respect to the notations of \cite{Tolstoy} we have
$X_\pm = L_\mp$, $X_0 = -2 L_0$ and $Q_\pm = \mp 2 i R_\mp$.)
These transformations satisfy the commutation/anticommutation relations
\begin{subeqnarray}
   [X_0, X_\pm] \,=\, \pm 2 X_\pm
        \quad   &  & \quad
   [X_+, X_-] \,=\, X_0
      \slabel{eq:sp2} \\[1mm]
   \{ Q_\pm, Q_\pm \} \,=\, \pm 2 X_\pm
        \quad   &  & \quad
   \{ Q_+, Q_- \} \,=\,  X_0
      \slabel{eq:osp12a} \\[1mm]
   [X_0, Q_\pm] \,=\, \pm Q_\pm \quad\qquad
      [X_\pm, Q_\pm] &\!=\!&  0  \quad\qquad
      [X_\pm, Q_\mp] \,=\,  -Q_\pm 
     \slabel{eq:osp12c}
\end{subeqnarray}
Note in particular that $X_\pm = Q_\pm^2$ and $X_0 = Q_+ Q_- + Q_- Q_+$.
It follows that any element of the Grassmann algebra
that is annihilated by $Q_\pm$
is also annihilated by the entire $\osp(1|2)$ algebra.

Now let us consider a $\sigma$-model in which the superfields $\nv_i$
are constrained to lie on the unit supersphere in $\R^{1|2}$,
i.e.\ to satisfy the constraint
\begin{equation}
   \nv_i\cdot \nv_i \;\equiv\; \sigma_i^2 + 2\lambda \psibar_i \psi_i  \;=\; 1
\ef.
\end{equation}
We can {\em solve}\/ this constraint by writing
\begin{equation}
\sigma_i  \;=\;  \pm (1 - 2\lambda \psibar_i \psi_i)^{1/2}
          \;=\;  \pm (1 - \lambda \psibar_i \psi_i)
  \;,
 \label{def.sigmai}
\end{equation}
exploiting the fact that $\psi_i^2 = \psibar_i^2 = 0$.
Let us henceforth take only the $+$ sign in \reff{def.sigmai},
neglecting the other solution
(the role played by these neglected Ising variables will be considered
 in more detail elsewhere \cite{us_forests_ON}),
so that
\begin{equation}
\sigma_i  \;=\; 1 - \lambda \psibar_i \psi_i \;.
 \label{def.sigmai_bis}
\end{equation}
We then have a purely fermionic model with variables $\psi,\psibar$
in which the $\ssp(2)$ transformations continue to act as in \reff{def.sp2}
while the fermionic transformations act via the ``hidden'' supersymmetry
\begin{subeqnarray}
 \delta \psi_i & = &
   \lambda^{-1/2}\, \epsilon \,(1-\lambda \psibar_i \psi_i)    \\
 \delta \psibar_i & = &
    \lambda^{-1/2}\,\epsilonbar \,(1-\lambda \psibar_i \psi_i)
 \label{eq.supersym.psionly}
\end{subeqnarray}
All of these transformations leave invariant the scalar product
\begin{equation}
   \nv_i\cdot\nv_j
   \;=\,
   1 \,-\, \lambda (\psibar_i - \psibar_j) (\psi_i - \psi_j)
     \,+\, \lambda^2 \psibar_i \psi_i \psibar_j \psi_j
   \;.
\end{equation}
The generators $Q_\pm$ are now defined as
\begin{subeqnarray}
  Q_+  & = &
      \lambda^{-1/2} \sum_{i\in V}  (1 - \lambda \psibar_i \psi_i) \partial_i 
      \;=\;
      \lambda^{-1/2} \partial \,-\,
          \lambda^{1/2} \sum_{i\in V} \psibar_i \psi_i \partial_i
   \\[1mm]
  Q_-  & = &   \lambda^{-1/2} \sum_{i\in V}  (1 - \lambda \psibar_i \psi_i)
      \bar{\partial}_i 
      \;=\;
      \lambda^{-1/2} \bar{\partial} \,-\,
          \lambda^{1/2} \sum_{i\in V} \psibar_i \psi_i \bar{\partial}_i
 \label{eq:def_Q_bis}
\end{subeqnarray}
where we recall the notations
$\partial = \sum\limits_{i \in V} \partial_{i}$
and $\bar{\partial} = \sum\limits_{i \in V} \bar{\partial}_{i}$.


Let us now show that the polynomials $f_A^{(\lambda)}$
defined as in~\reff{eq:new_A} are $\OSP(1|2)$-invariant,
i.e.\ are annihilated by all elements of the $\osp(1|2)$ algebra.
As noted previously, it suffices to show that the $f_A^{(\lambda)}$
are annihilated by $Q_\pm$.
Applying the definitions \reff{eq:def_Q_bis}, we have
\begin{equation}
   Q_- \tau_A  \;=\;  \lambda^{-1/2} \bar{\partial} \tau_A
\end{equation}
and hence
\begin{equation}
   Q_+ Q_- \tau_A  \;=\;  \lambda^{-1} \partial \bar{\partial} \tau_A
                          \,-\, |A| \tau_A
   \;,
\end{equation}
so that 
\begin{equation}
   f_A^{(\lambda)} \;=\; \lambda (1 + Q_+ Q_-) \tau_A
   \;.
\label{eq:refer}
\end{equation}
The next step is to compute $Q_+ f_A^{(\lambda)}$:  since
\begin{eqnarray}
   & &
    Q_+ (1 + Q_+ Q_-)  \;=\; Q_+ + Q_+^2 Q_-
                       \;=\; Q_+ + X_+ Q_-
         \nonumber \\
   & & \qquad           \;=\; Q_+ + [X_+, Q_-] + Q_- X_+
                       \;=\; Q_+ - Q_+ + Q_- X_+
                       \;=\; Q_- X_+
     \qquad
\end{eqnarray}
by the relations \reff{eq:osp12a}/\reff{eq:osp12c},
while it is obvious that $X_+ \tau_A = 0$,
we conclude that $Q_+ f_A^{(\lambda)} = 0$,
i.e.\ $f_A^{(\lambda)}$ is invariant under the transformation $Q_+$.
A similar calculation of course works for $Q_-$.\footnote{
   We are grateful to an anonymous referee for suggesting this proof.
   An alternate proof that $Q_\pm f_A^{(\lambda)} = 0$,
   based on direct calculation using the definition \reff{eq.deff_A}
   of $f_A^{(\lambda)}$, can be found in the first preprint version
   of this article ({\tt arXiv:0706.1509v1}):
   see equations (7.8)--(7.11) there.
}

In fact, the $\OSP(1|2)$-invariance of $f_A^{(\lambda)}$
can be proven in a simpler way by writing $f_A^{(\lambda)}$
explicitly in terms of the scalar products $\nv_i \cdot \nv_j$ for $i,j \in A$.
Note first that
\begin{subeqnarray}
    f^{(\lambda)}_{\{i,j\}}
    & = &
    - \lambda \psibar_i\psi_i\psibar_j\psi_j
     \,+\,  (\psibar_i - \psibar_j) (\psi_i - \psi_j)   \\[2mm]
    & = &
    \frac{1}{\lambda} \left( 1 - \nv_i \cdot \nv_j \right)
         \slabel{eq.fijlambda.ninj.b}  \\[2mm]
    & = &
    \frac{(\nv_i-\nv_j)^2}{2\lambda}
   \;.
 \label{eq.fijlambda.ninj}
\end{subeqnarray}
By Corollary~\ref{corol:nil2}, we obtain
\begin{subeqnarray}
   f^{(\lambda)}_{\{i_1,i_2,\ldots,i_{k}\}}
   & = &
   \frac{1}{\lambda^{k-1}} \, (1 - \nv_{i_1}\cdot\nv_{i_2})
        \, (1 - \nv_{i_2}\cdot\nv_{i_3}) \,\cdots\,
           (1 - \nv_{i_{k-1}}\cdot\nv_{i_k})
        \qquad \\[2mm]
   & = &
   \frac{1}{(2\lambda)^{k-1}} \, (\nv_{i_1}-\nv_{i_2})^2 
        \, (\nv_{i_2}-\nv_{i_3})^2 \,\cdots\, (\nv_{i_{k-1}}-\nv_{i_k})^2
\;.
 \slabel{eq.fAlambda.ninj.b}
 \label{eq.fAlambda.ninj}
\end{subeqnarray}
Note the striking fact that the right-hand side of \reff{eq.fAlambda.ninj}
is invariant under all permutations of $i_1,\ldots,i_k$,
though this fact is not obvious from the formulae given,
and is indeed false for vectors in Euclidean space $\R^N$ with $N \neq -1$.
Moreover, the path $i_1,\ldots,i_k$ that is implicit in the right-hand side
of \reff{eq.fAlambda.ninj} could be replaced by {\em any tree}\/
on the vertex set $\{i_1,\ldots,i_k\}$,
and the result would again be the same (by Corollary~\ref{corol:nil2}).

It follows from \reff{eq.fijlambda.ninj}/\reff{eq.fAlambda.ninj}
that the subalgebra generated by the scalar products
$\nv_i \cdot \nv_j$ for $i,j \in V$
is identical with the subalgebra generated by the $f_A^{(\lambda)}$
for $A \subseteq V$, for any $\lambda \neq 0$.
Therefore, the {\em most general}\/ $\OSP(1|2)$-symmetric Hamiltonian
depending on the $\{\nv_i\}_{i \in V}$
is precisely the one discussed in the Remark
at the end of Section~\ref{sec:integrals},
namely in which the action contains all possible products
$f_\scrc^{(\lambda)} = \prod\limits_\alpha f_{C_\alpha}^{(\lambda)}$,
where $\{C_\alpha\}$ is a partition of $V$.

In Appendix~\ref{app:grass.2} we will prove a beautiful alternative formula
for $f^{(\lambda)}_{\{i_1,i_2,\ldots,i_{k}\}}$:
\begin{equation}
   f^{(\lambda)}_{\{i_1,i_2,\ldots,i_{k}\}}
   \;=\;
   \frac{1}{k! \, \lambda^{k-1}} \det M
 \label{eq.fA.detformula}
\end{equation}
where $M$ is the $k \times k$ matrix of scalar products
$M_{rs} = \nv_{i_r} \cdot \nv_{i_s}$.
In this formula, unlike \reff{eq.fAlambda.ninj},
the symmetry under all permutations of $i_1,\ldots,i_k$ is manifest.
We remark that the determinant of a matrix of inner products
is commonly called a \emph{Gram determinant}\/ \cite[p.~110]{Lancaster_85}.

Finally, we need to consider the behavior of the integration measure
in \reff{eq.Zgrass}, namely
\begin{equation}
  \mathcal{D}_{V,\bt}(\psi, \psibar)
  \;=\;
  \prod_{i \in V} \dx{\psi_i} \, \dx{\psibar_i} \, e^{ t_i \psibar_i \psi_i}
  \;,
\end{equation}
under the supersymmetry \reff{eq.supersym.psionly}.
In general this measure is {\em not}\/ invariant under
\reff{eq.supersym.psionly},
but in the special case $t_i = \lambda$ for all $i$,
it {\em is}\/ invariant, in the sense that
\begin{equation}
  \int \mathcal{D}_{V,\lambda}(\psi, \psibar)  \: \delta F(\psi,\psibar)
  \;=\;
  0
\end{equation}
for any function $F(\psi,\psibar)$.
Indeed, $\mathcal{D}_{V,\lambda}(\psi, \psibar)$ is invariant
more generally under {\em local}\/ supersymmetry transformations in which
separate generators $\epsilon_i,\epsilonbar_i$
are used at each vertex $i$.
To see this, let us focus on one site $i$ and write
$F(\psi,\psibar) = a + b \psi_i + c \psibar_i + d \psibar_i \psi_i$
where $a,b,c,d$ are polynomials in the $\{\psi_j,\psibar_j\}_{j \neq i}$
(which may contain both Grassmann-even and Grassmann-odd terms).
Then
\begin{subeqnarray}
   \delta F
   & = &
   \lambda^{1/2}\, \left[ b \epsilon_i \sigma_i \,+\,
   c \epsilonbar_i \sigma_i \,+\,
   d \left( \epsilonbar_i \sigma_i \psi_i + \psibar_i \epsilon_i \sigma_i \right) \right]
          \\[2mm]
   & = &
   \sigma_i \,\lambda^{1/2}\,\biggl[ b \epsilon_i \,+\, c \epsilonbar_i \,+\,
                    d (\epsilonbar_i \psi_i + \psibar_i \epsilon_i) \biggr]
   \;.
\end{subeqnarray}
Since $\sigma_i = e^{-\lambda \psibar_i \psi_i}$,
this cancels the factor $e^{t_i \psibar_i \psi_i}$
from the measure (since $t_i = \lambda$)
and the integral over $d\psi_i \, d\psibar_i$ is zero
(because there are no $\psibar_i \psi_i$ monomials).
Thus, the measure $\mathcal{D}_{V,\bt}(\psi, \psibar)$
is invariant under the local supersymmetry at site $i$
whenever $t_i = \lambda$.
If this occurs for all $i$, then the measure is invariant under
the global supersymmetry \reff{def.supersym}.

The $\OSP(1|2)$-invariance of $\mathcal{D}_{V,\lambda}(\psi, \psibar)$
can be seen more easily by writing the manifestly invariant combination
\begin{subeqnarray}
   \delta(\nv_i^2 -1) \, d\nv_i
   & = &
   \delta (\sigma_i^2 + 2 \lambda \psibar_i \psi_i -1) \,
       d\sigma_i \, d\psi_i \, d\psibar_i
       \\[2mm]
   & = &
   e^{\lambda \psibar_i \psi_i} \,
      \delta \big(\sigma_i - (1 -  \lambda \psibar_i \psi_i) \big) \,
       d\sigma_i \, d\psi_i \, d\psibar_i
   \;,
 \label{eq.measure}
 \slabel{eq.measure.b}
\end{subeqnarray}
where the factor $e^{\lambda \psibar_i \psi_i}$
comes from the inverse Jacobian.
Integrating out $\sigma_i$ from \reff{eq.measure.b},
we obtain $e^{\lambda \psibar_i \psi_i} \, d\psi_i \, d\psibar_i$.

As a consequence of \reff{eq.fAlambda.ninj.b} and \reff{eq.measure.b},
the generating function \reff{cor1.Zgrass.k-uniform}
for spanning hyperforests in a $k$-uniform hypergraph can be rewritten as
\begin{eqnarray}
   & & \!\!\!\!\!
   \sum_{F\in \scrf(G)}
      \!
        \Biggl( \prod\limits_{A \in F} w_A \! \Biggr)
      \; \lambda^{k(F)}
   \;\,=\;\,
   \int \Biggl( \prod_{i\in V} \delta(\nv_i^2 -1) \, d\nv_i \Biggr)
    \;\times
         \nonumber \\
   & &
\,\exp\!\left[  \frac{1}{(2\lambda)^{k-1}} \sum_{i_1,\ldots,i_k\in V}
                \frac{L_{i_1,\ldots,i_k}}{(k-2)!} (\nv_{i_1}-\nv_{i_2})^2 
                (\nv_{i_2}-\nv_{i_3})^2 \cdots (\nv_{i_{k-1}}-\nv_{i_{k}})^2
        \right] 
   \,. \qquad
\end{eqnarray}
In the special case $k=2$, this result appears in \cite{us_prl}.

\section{Conclusions}
   \label{sec:concl}

In this paper we have applied techniques of Grassmann algebra,
first used in \cite{us_prl} to obtain the generating function of spanning
forests in a graph --- generalizing Kirchhoff's matrix-tree theorem ---
to a wider class of models associated to hypergraphs.
The key role in our analysis is played by a set of
simple algebraic rules (Lemma~\ref{lem.fAfB})
that express, in a certain sense, the fermionic-bosonic cancellation
associated to the underlying $\OSP(1|2)$ supersymmetry.
This algebraic approach allows for notably simplified proofs
and for strong generalizations.

In particular, we are able to obtain combinatorial interpretations
in terms of spanning hyperforests
for the partition function (Section~\ref{sec:integrals})
and correlation functions (Section~\ref{sec:corrfn})
of a fairly general class of fermionic models living on hypergraphs.
In that subset of the results where the $\OSP(1|2)$ supersymmetry is preserved,
the combinatorial weights of the hyperforest configurations become
(perhaps not surprisingly) notably simpler.
Among other things, we obtain the generating function of
unrooted spanning hyperforests on a weighted hypergraph
(together with a family of relevant combinatorial observables)
as an $\OSP(1|2)$-invariant fermionic integral.



Finally, in Appendix~\ref{app:andrea}
we present a graphical formalism
for proving both the classical matrix-tree theorem
and numerous extensions thereof,
which can serve as an alternative to the algebraic approach
used in the main body of this paper
and which we hope will have further applications.

In a follow-up paper \cite{CSS:subalgebra}
we shall study in more detail the Grassmann subalgebra
that is generated by the elements $f_A^{(\lambda)}$
as $A$ ranges over all nonempty subsets of $V$.

It is also natural to ask about extensions of this work in which
combinatorial interpretations are obtained for
statistical-mechanical models with other supersymmetry groups.
We are currently studying models with $\OSP(1|2n)$ and $\OSP(2|2)$
supersymmetries and hope to report the results in the near future.

\appendix

\section{A determinantal formula for $f_A^{(\lambda)}$}
   \label{app:grass.2}

The main purpose of this appendix is to prove the
determinantal formula \reff{eq.fA.detformula} for $f_A^{(\lambda)}$.
Along the way we will obtain a rather more general
graphical representation of certain determinants.

Let $A = (a_{ij})_{i,j=1}^n$ be a matrix whose elements belong to
a commutative ring $R$.
The determinant is defined as usual by
\begin{equation}
   \det A  \;=\;
   \sum_{\pi \in \Pi_n}
      \sgn(\pi) \prod_{i=1}^n a_{i \pi(i)}
   \;,
 \label{eq.def_det}
\end{equation}
where the sum runs over permutations $\pi$ of $[n] := \{1,\ldots,n\}$,
and 

$\sgn(\pi) =(-1)^{\#(\hbox{\scriptsize even cycles of }\pi)}$
is the sign of the permutation $\pi$.

We begin with a formula for the determinant of the sum of two matrices
in terms of minors, which ought to be well known
but apparently is not\footnote{
   This formula can be found in
   \cite[pp.~162--163, Exercise 6]{Marcus_75}
   and \cite[pp.~221--223]{Korepin_93}.
   It can also be found
   --- albeit in an ugly notation that obscures what is going on ---
   in
   \cite[pp.~145--146 and 163--164]{Marcus_75}
   \cite[pp.~31--33]{Rump_97}
   \cite[pp.~281--282]{Prells_03};
   and in an even more obscure notation in
   \cite[p.~102, item~5]{Aitken_48}.
   We remark that an analogous formula holds (with the same proof)
   in which all three occurrences of determinant are replaced by
   permanent and the factor $\epsilon(I,J)$ is omitted.
}:


\begin{lemma}
   \label{lemma.detA+B}
Let $A$ and $B$ be $n \times n$ matrices whose elements belong to
a commutative ring $R$.  Then\footnote{
   The determinant of an empty matrix is of course defined to be 1.
   This makes sense in the present context even if the ring $R$
   lacks an identity element:  the term $I=J=\emptyset$ contributes
   $\det B$ to the sum \reff{eq.lemma.detA+B}, while the term
   $I=J=[n]$ contributes $\det A$.
}
\begin{equation}
\det(A+B) \;=\; \sum_{\begin{scarray}
                             I,J \subseteq [n] \\
                            |I| = |J|
                          \end{scarray}}
                      \epsilon(I,J) \,  (\det A_{IJ})  (\det B_{I^c J^c})
   \;,
 \label{eq.lemma.detA+B}
\end{equation}
where $\epsilon(I,J) = (-1)^{\sum_{i \in I} i + \sum_{j \in J} j}$
is the sign of the permutation that takes $II^c$ into $JJ^c$
(where the sets $I,I^c,J,J^c$ are all written in increasing order).
\end{lemma}

\proof
Using the definition of determinant and expanding the products, we have
\begin{equation}
   \det(A+B)  \;=\;
   \sum_{\pi \in \Pi_n} \sgn(\pi) 
   \sum_{I \subseteq [n]} \prod_{i \in I} a_{i \pi(i)}
                          \prod_{\ell \in I^c} b_{\ell \pi(\ell)}
   \;.
\end{equation}
Define now $J = \pi[I]$.  Then we can interchange the order of summation:
\begin{equation}
   \det(A+B)  \;=\;
   \sum_{\begin{scarray}
            I,J \subseteq [n] \\
            |I| = |J|
         \end{scarray}}
   \sum_{\begin{scarray}
            \pi \in \Pi_n \\
            \pi[I] = J
         \end{scarray}}
   \sgn(\pi) 
   \prod_{i \in I} a_{i \pi(i)}
   \prod_{\ell \in I^c} b_{\ell \pi(\ell)}
   \;.
\end{equation}
Suppose now that $|I|=|J|=k$, and let us write
$I = \{i_1,\ldots,i_k\}$ and $J = \{j_1,\ldots,j_k\}$
where the elements are written in increasing order,
and likewise
$I^c = \{\ell_1,\ldots,\ell_{n-k}\}$ and $J = \{m_1,\ldots,m_{n-k}\}$.
Let $\pi' \in \Pi_k$ and $\pi'' \in \Pi_{n-k}$ be the permutations
defined so that
\begin{subeqnarray}
   \pi(i_\alpha) = j_\beta  & \longleftrightarrow &  \pi'(\alpha) = \beta \\
   \pi(\ell_\alpha) = m_\beta  & \longleftrightarrow &  \pi''(\alpha) = \beta
\end{subeqnarray}
It is easy to see that $\sgn(\pi) = \sgn(\pi') \sgn(\pi'') \epsilon(I,J)$.
The formula then follows by using twice again the definition of determinant.
\qed

\begin{corollary}
   \label{cor.det.rank}
Let $A$ and $B$ be $n \times n$ matrices whose elements belong to
a commutative ring $R$.
Then $\det(A + \lambda B)$ is a polynomial in $\lambda$
of degree at most ${\rm rank}(B)$,
where ``rank'' here means determinantal rank
(i.e.\ the order of the largest nonvanishing minor).
\end{corollary}

\proof
This is an immediate consequence of the formula \reff{eq.lemma.detA+B},
since all minors of $B$ of size larger than its rank vanish by definition.
\qed

Next recall the traditional graphical representation of the determinant:

\begin{lemma}
   \label{lemma.det.1}
Let $C = (c_{ij})_{i,j=1}^n$ be a matrix whose elements belong to
a commutative ring $R$.  Then
\begin{equation}
   \det(-C)  \;=\;
   \sum_{\vec{G}}  (-1)^{\#(\hbox{\scriptsize\rm cycles of } \vec{G})}
   \prod_{ij \in E(\vec{G})} c_{ij}
   \;,
 \label{eq.lemma.det.1}
\end{equation}
where the sum runs over all permutation digraphs $\vec{G}$
on the vertex set $\{1,2,\ldots,n\}$,
i.e.\ all directed graphs in which each connected component
is a directed cycle (possibly of length 1).
\end{lemma}

\proof
This is an immediate consequence of \reff{eq.def_det}
and the fact that
$(-1)^{\#(\hbox{\scriptsize even cycles of }\pi)}$ $=
 (-1)^{\#(\hbox{\scriptsize cycles of }\pi)}
 (-1)^{\#(\hbox{\scriptsize odd cycles of }\pi)}$.
\qed

Now let ${\bf a} = (a_i)_{i=1}^n$ and ${\bf b} = (b_i)_{i=1}^n$
be a pair of vectors with elements in the ring $R$.
The main result of this appendix is
the following generalization of Lemma~\ref{lemma.det.1}:

\begin{lemma}
   \label{lemma.det.2}
Let $C = (c_{ij})_{i,j=1}^n$ be a matrix whose elements belong to
a commutative ring $R$,
and let ${\bf a} = (a_i)_{i=1}^n$ and ${\bf b} = (b_i)_{i=1}^n$
be vectors with elements in $R$.  Then
\begin{equation}
   \det({\bf a} {\bf b}^{\rm T} - C)  \;=\;
   \det(-C) \,+\,
   \sum_{\vec{G}}  (-1)^{\#(\hbox{\scriptsize\rm cycles of } \vec{G})}  \,
   b_{s(\vec{G})} \, a_{t(\vec{G})} \prod_{ij \in E(\vec{G})} c_{ij}
   \;,
 \label{eq.lemma.det.2}
\end{equation}
where the sum runs over all directed graphs $\vec{G}$ on the vertex set
$\{1,2,\ldots,n\}$ in which one connected component is a
directed path (possibly of length 0, i.e.\ an isolated vertex)
from source $s(\vec{G})$ to sink $t(\vec{G})$
and all the other connected components are directed cycles
(possibly of length 1).
\end{lemma}

\proof
Introduce an indeterminate $\lambda$ and let us compute
$\det(\lambda {\bf a} {\bf b}^{\rm T} - C)$,
working in the polynomial ring $R[\lambda]$,
by substituting $c_{ij} - \lambda a_i b_j$ in place of $c_{ij}$
in \reff{eq.lemma.det.1}.
The term of order $\lambda^0$ is $\det(-C)$,
which is given by \reff{eq.lemma.det.1}.
In the term of order $\lambda^1$, one edge $ij$ in $\vec{G}$
carries a factor $-a_i b_j$ and the rest carry matrix elements of $C$.
Setting $\vec{G}' = \vec{G} \setminus ij$,
we see that $\vec{G}'$ has one less cycle than $\vec{G}$
[thereby cancelling the minus sign]
and has a path running from source $s(\vec{G}') = j$
to sink $t(\vec{G}') = i$.
Dropping the prime gives \reff{eq.lemma.det.2}.
Terms of order $\lambda^2$ and higher vanish
by Corollary~\ref{cor.det.rank}
because ${\bf a} {\bf b}^{\rm T}$ has rank 1.
\qed

\begin{corollary}
   \label{cor.det.2a}
Let $C = (c_{ij})_{i,j=1}^n$ be a matrix whose elements belong to
a commutative ring-with-identity element $R$,
and let $E$ be the $n \times n$ matrix with all elements $1$.  Then
\begin{equation}
   \det(E - C)  \;=\;
   \det(-C) \,+\,
   \sum_{\vec{G}}  (-1)^{\#(\hbox{\scriptsize\rm cycles of } \vec{G})}
   \prod_{ij \in E(\vec{G})} c_{ij}
   \;,
 \label{eq.cor.det.2a}
\end{equation}
where the sum runs over all directed graphs $\vec{G}$ on the vertex set
$\{1,2,\ldots,n\}$ in which one connected component is a
directed path (possibly of length 0, i.e.\ an isolated vertex)
and all the other connected components are directed cycles
(possibly of length 1).
\end{corollary}


The following result is an immediate consequence of
Lemma~\ref{lemma.det.1} and Corollary~\ref{cor.det.2a}:

\begin{corollary}
   \label{cor.det.1}
Let $C = (c_{ij})_{i,j=1}^n$ be a matrix whose elements belong to
a commutative ring-with-identity-element $R$ and satisfy
$c_{i_1 i_2} c_{i_2 i_3} \cdots c_{i_{k-1} i_k} c_{i_k i_1} = 0$
for all $i_1,\ldots,i_k$  $(k \ge 1)$,
and let $E$ be the $n \times n$ matrix with all elements $1$.  Then
\begin{equation}
   \det(E-C)  \;=\;
   \sum_{\vec{P}} 
   \prod_{ij \in E(\vec{P})} c_{ij}
   \;,
 \label{eq.cor.det.1}
\end{equation}
where the sum runs over all directed paths $\vec{P}$ on the vertex set
$\{1,2,\ldots,n\}$.
(There are $n!$ such contributions.)
\end{corollary}

\proof
The hypotheses on $C$ lead to the vanishing of all terms
containing at least one cycle (including cycles of length 1).
Therefore, the only remaining possibility is a single directed path.
\qed

Let us now specialize Corollary~\ref{cor.det.1}
to the case in which the commutative ring $R$
is the even subalgebra of our Grassmann algebra,
and the matrix $C$ is given by
\begin{align}
   c_{ii}                & =  0       \\[1mm]
   c_{ij} = c_{ji}   & =  \lambda f_{\{i,j\}}^{(\lambda)}
                                   \qquad \hbox{for }  i \neq j
\end{align}
The hypothesis $c_{i_1 i_2} c_{i_2 i_3} \cdots c_{i_{k-1} i_k} c_{i_k i_1} = 0$
is an immediate consequence of Corollary~\ref{corol:nil2}.
Moreover, by equation~\reff{eq.fijlambda.ninj.b} we have
$(E-C)_{ij} = \nv_i \cdot \nv_j$.
In the expansion \reff{eq.cor.det.1}
we obtain $n!$ terms, each of which is of the form
$\lambda^{n-1}$ times $\prod_{ij \in E(\vec{P})} f_{\{i,j\}}^{(\lambda)}$
for some directed path $\vec{P}$ on the vertex set $\{1,2,\ldots,n\}$.
But by Corollary~\ref{corol:nil2},
each such product equals $f_{\{1,\ldots,n\}}^{(\lambda)}$,
so this proves the determinantal formula \reff{eq.fA.detformula}
for $f_{\{1,\ldots,n\}}^{(\lambda)}$.

\section{Graphical proof of some generalized matrix-tree theorems}
   \label{app:andrea}




%
%
%
%

In this appendix we shall give a ``graphical'' proof
of the classical matrix-tree theorem as well as
a number of extensions thereof,
by interpreting in a graphical way the terms of a
formal Taylor expansion of an action belonging to the even subalgebra of a
Grassmann algebra.
(We require the action to belong to the even subalgebra
 in order to avoid ordering ambiguities
 when exponentiating a sum of terms.)
Some of these extensions of the matrix-tree theorem are already set forth
in the main body of this paper,
where they are proven by an ``algebraic'' method
based on Lemma~\ref{lem.fAfB} and its corollaries.
Other more exotic extensions are described here with an eye to future work;
they could also be proven by suitable variants of the algebraic technique.

Curiously enough, it turns out that the more general is the fact
we want to prove, the {\em easier}\/ is the proof;
indeed, the most general facts ultimately become almost tautologies
on the rules of Grassmann algebra and integration.
The only extra feature of the most general facts
is that the ``zoo'' of graphical combinatorial objects
has to become wider (and wilder).

So, in this exposition we shall start by describing the most general situation,
and then show how, when special cases are chosen for the
parameters in the action, a corresponding simplification
occurs also in the combinatorial interpretation.

\subsection{General result}


Consider a hypergraph $G=(V,E)$ as defined in Section~\ref{sec:graph},
i.e.\ $V$ is a finite set and $E$ is a set of subsets of $V$,
each of cardinality at least 2, called {\em hyperedges}\/.
As usual we introduce a pair $\psi_i,\psibar_i$
of Grassmann generators for each $i \in V$.
We shall consider actions of the form
\begin{equation}
   \mathcal{S}(\psi, \psibar)
   \;=\;
   \sum_{A \in E} \mathcal{S}_A(\psi, \psibar)
   \;,
\end{equation}
where
\begin{equation}
   \mathcal{S}_A(\psi, \psibar) 
   \;=\;
   w_A^* \tau_A  \,+\,  \sum_{i \in A} w_{A;i} \, \tau_{A \setminus i}
     \,+\!\!  \sum_{\begin{scarray}
                      i,j \in A \\
                      i \neq j
                  \end{scarray}}
            \! w_{A;ij} \, \psi_i \psibar_j \tau_{A \setminus \{i,j\} }
 \label{def.SA}
\end{equation}
and $\tau_A = \prod_{i \in A} \psibar_i \psi_i$.
Please note that the form \reff{def.SA} resembles
the definition \reff{eq.deff_A} of $f_A^{(\lambda)}$
the same monomials appear, but now each one is multiplied by an independent
indeterminate.
Thus, for each hyperedge $A$ of cardinality $k$ we have $k^2+1$ parameters:
$w_A^*$, $\{ w_{A;i} \}_{i \in A}$ and 
$\{ w_{A;i,j} \}_{(i \neq j) \in A}$.
[We have chosen, for future convenience, to write the last term
 in \reff{eq.deff_A} as $+\psi_i \psibar_j$ rather than $-\psibar_i \psi_j$.]

Please note that, for $|A| > 2$, all pairs of terms in 
$\mathcal{S}_A(\psi, \psibar)$ have a vanishing product,
because they contain at least $2(2|A|-2) = 4|A|-4$ fermions
in a subalgebra (over $A$) that has only $2|A|$ distinct fermions.
As a consequence, we have in this case
\begin{equation}
\exp[\mathcal{S}_A(\psi, \psibar)]  \;=\;
1+ \mathcal{S}_A(\psi, \psibar) 
\ef.
\end{equation}
On the other hand, if $|A|=2$ (say, $A=\{i,j\}$),
we have two nonvanishing cross-terms:
\begin{subeqnarray}
(w_{A;i} \, \psibar_j \psi_j)  \,
(w_{A;j} \, \psibar_i \psi_i) 
  & = & w_{A;i} w_{A;j} \, \psibar_i \psi_i \psibar_j \psi_j
        \\[2mm]
(w_{A;ij} \, \psi_i \psibar_j) \,
(w_{A;ji} \, \psi_j \psibar_i)
  & = & -w_{A;ij} w_{A;ji} \, \psibar_i \psi_i \psibar_j \psi_j
\end{subeqnarray}
where the minus sign comes from commutation of fermionic fields.
So we can write in the general case
\begin{equation}
\exp[\mathcal{S}_A(\psi, \psibar)] \;=\;
1+ \widehat{\mathcal{S}}_A(\psi, \psibar) 
  \;,
\end{equation}
where $\widehat{\mathcal{S}}_A(\psi, \psibar)$ is defined like
$\mathcal{S}_A(\psi, \psibar)$ but with the parameter $w_A^*$ replaced by
\begin{equation}
\widehat{w}_A^*  \;=\;
   \begin{cases}
       w_A^* + w_{A;i} w_{A;j} - w_{A;ij} w_{A;ji}
                 & \hbox{if }  A=\{i,j\}   \\[1mm]
       w_A^*     & \hbox{if } |A| \geq 3
   \end{cases}
\end{equation}

Consider now a Grassmann integral of the form
\begin{equation}
   \int \! \mathcal{D}(\psi, \psibar) \, \scro_{I,J} \,
  \exp \Biggl[ \sum_i t_i \psibar_i \psi_i
               \,+\,  \sum_{A \in E} \mathcal{S}_A(\psi, \psibar) 
       \Biggr]
   \;, \quad
 \label{eq.A1}
\end{equation}
where $\bt = (t_i)_{i \in V}$ are parameters,
$I=(i_1,i_2,\dots, i_k) \in V^k$ and $J=(j_1,j_2,\dots, j_k) \in V^k$
are ordered $k$-tuples of vertices, and
\begin{equation}
\mathcal{O}_{I,J}
  \; := \;
\psibar_{i_1} \psi_{j_1} \cdots
\psibar_{i_k} \psi_{j_k}
\end{equation}
[cf.\ \reff{eq.operO}].
Here the $i_1,\dots, i_k$ must be all distinct, as must the $j_1,\dots, j_k$,
but there can be overlaps between the sets
${\sf I} = \{i_1,i_2,\dots, i_k\}$
and ${\sf J} = \{j_1,j_2,\dots, j_k\}$.\footnote{
   Please note the distinction between the
   {\em ordered}\/ $k$-tuple $I=(i_1,i_2,\dots, i_k)$,
   here written in italic font,
   and the {\em unordered}\/ set ${\sf I} = \{i_1,i_2,\dots, i_k\}$,
   here written in sans-serif font.
}
We intend to show that \reff{eq.A1} can be interpreted combinatorially
as a generating function for rooted oriented\footnote{
   We shall define later what we mean by ``orienting'' a hyperedge $A$:
   it will correspond to selecting a single vertex $i \in A$
   as the ``outgoing'' vertex.
}
spanning sub(hyper)graphs of $G$,
in which each connected component is either a (hyper-)tree
or a (hyper-)unicyclic.
In the case of a unicyclic component, the rest of the component
is oriented towards the cycle,
and no vertex from ${\sf I} \cup {\sf J}$ lies in the component.
In the case of a tree component, either
(a) no vertex from ${\sf I} \cup {\sf J}$ is in the component, and then
there is either a special ``root'' vertex or a ``root'' hyperedge,
all the rest of the tree being oriented towards it,
or (b) the component contains a single vertex from ${\sf I} \cap {\sf J}$,
which is the root vertex, and the tree is again oriented towards it,
or (c) the component contains exactly one vertex from ${\sf I}$
and one from ${\sf J}$, a special oriented path connecting them, and
all the rest is oriented towards the path.
The weight of each configuration is essentially
the product of $t_i$ for each root $i \notin {\sf I} \cup {\sf J}$
and an appropriate weight ($\widehat{w}_A^*$, $w_{A;i}$ or $w_{A;ij}$)
for each occupied hyperedge,
along with a $-$ sign for each unicyclic using the $w_{A;ij}$'s
and a single extra $\pm$ sign corresponding to
the pairing of vertices of ${\sf I}$ to vertices of ${\sf J}$
induced by being in the same component.
(This same sign appeared already in Section~\ref{sec:corrfn}.)

Kirchhoff's matrix-tree theorem arises when all the hyperedges $A$
have cardinality 2 (i.e.\ $G$ is an ordinary graph),
${\sf I} = {\sf J} = \{i_0\}$ for some vertex $i_0$,
all $t_i=0$, all $w_A^*=0$, and $w_{A;i}=w_{A;ij}=w_A$.
The principal-minors matrix-tree theorem is obtained by allowing
${\sf I} = {\sf J}$ of arbitrary cardinality $k$,
while the all-minors matrix-tree theorem is obtained by allowing also
${\sf I} \neq {\sf J}$.
Rooted forests with root weights $t_i$ can be obtained
by allowing $t_i \neq 0$.
On the other hand, unrooted forests are obtained by taking all $t_i=\lambda$,
${\sf I} = {\sf J} = \emptyset$, $w_A^*=-\lambda w_A$ and the rest as above.
[More generally, unrooted hyperforests are obtained by taking all $t_i=\lambda$,
${\sf I} = {\sf J} = \emptyset$, $w_A^*=-\lambda (|A|-1) w_A$
and the rest as above.]
The sequences $I$ and $J$ are used mainly in order to obtain
expectation values of certain connectivity patterns
in the relevant ensemble of spanning subgraphs.

Let us now prove all these statements, and give precise expressions for the
weights of the configurations, which until now have been left deliberately
vague in order not to overwhelm the reader.

We start by manipulating \reff{eq.A1}, exponentiating the action to obtain
\begin{equation}
\label{eq.A1b}
   \int \! \mathcal{D}(\psi, \psibar) \,  \scro_{I,J} \,
\biggl( \prod_{i \in V} (1 + t_i \psibar_i \psi_i) \biggr)
\biggl( \prod_{A \in E} (1 + \widehat{\mathcal{S}}_A ) \biggr)
\end{equation}
or, expanding the last products,
\begin{equation}
\label{eq.A1b2}
\sum_{\substack{
          V' \subseteq V \setminus ({\sf I} \cup {\sf J}) \\[1mm]
          E' \subseteq E
      }}
\biggl( \prod_{i \in V'} t_i \biggr)
   \int \! \mathcal{D}(\psi, \psibar) \, \scro_{I \cup V',J \cup V'} \,
\biggl( \prod_{A \in E'} \widehat{\mathcal{S}}_A \biggr)
   \;,
\end{equation}
where $I \cup V'$ consists of the sequence $I$ followed by
the list of elements of $V'$ in any chosen order,
and $J \cup V'$ consists of the sequence $J$ followed by
the list of elements of $V'$ in the {\em same}\/ order.

We now give a graphical representation and a fancy name to each kind of
monomial in the expansion \reff{eq.A1b2},
as shown in Table~\ref{table_graphical}.
%
%
%
Please note that in this graphical representation
a solid circle $\bullet$ corresponds to a factor $\psibar_i \psi_i$,
an open circle $\circ$ corresponds to a factor $\psibar_i$,
and a cross $\times$ corresponds to a factor $\psi_i$.

\begin{table}[tp]
\begin{center}
\begin{tabular}{|l|c|p{3.6cm}|}
\hline
\multicolumn{3}{|c|}{Factors coming from 
 $\mathcal{O}_{I \cup V',J \cup V'}$%
\raisebox{-6pt}{\rule{0pt}{19pt}}
}
\\
\hline
$\psibar_i \psi_i$
&
\raisebox{-8pt}{%
\setlength{\unitlength}{10pt}
\begin{picture}(0,0)
\put(1.6,1){$i$}
\end{picture}%
\includegraphics[scale=1., bb=15 200 65 220, clip=true]
  {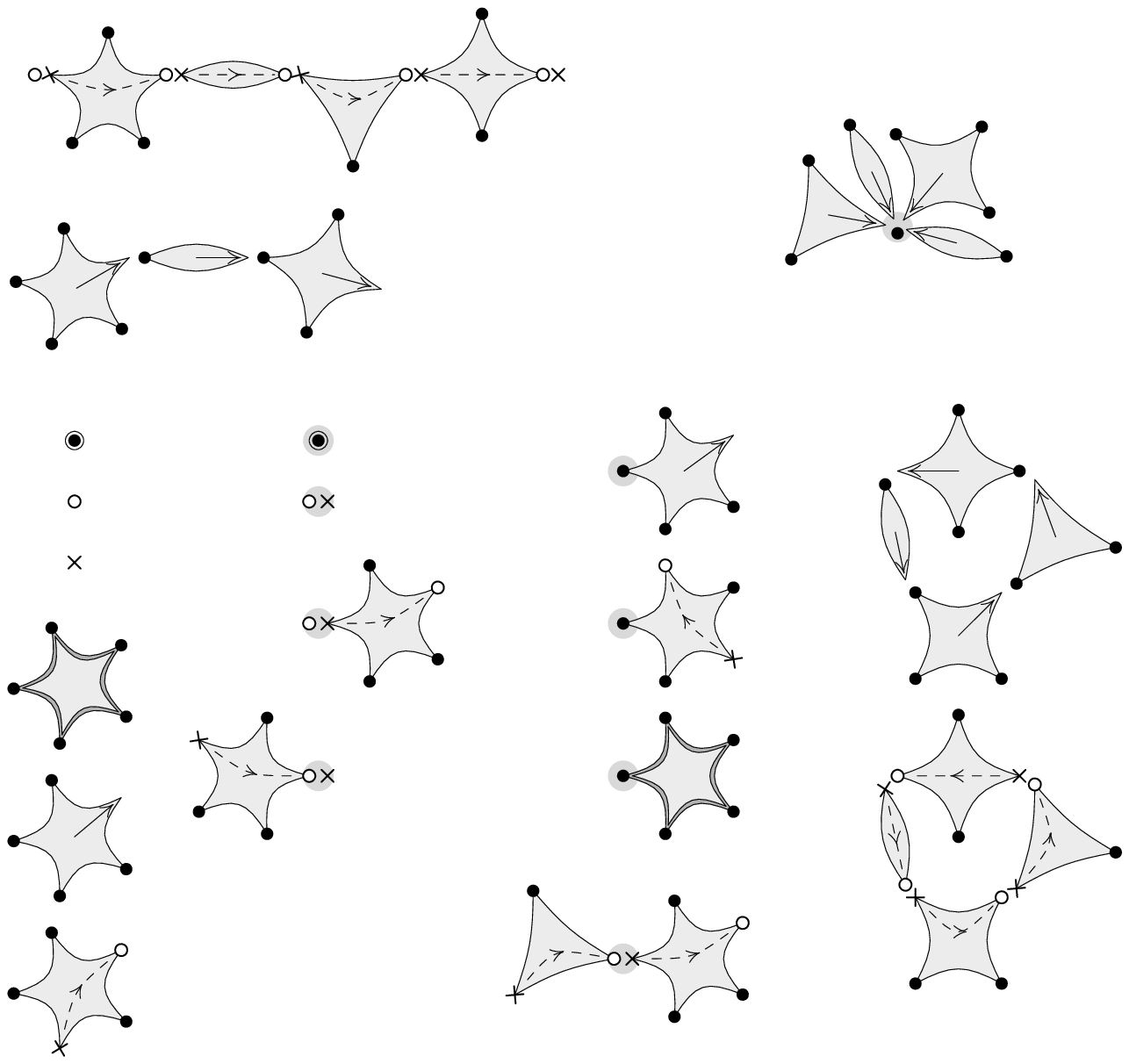}}
&
root vertex
\\
$\psibar_i$
&
\raisebox{-8pt}{%
\setlength{\unitlength}{10pt}
\begin{picture}(0,0)
\put(1.6,1){$i$}
\end{picture}%
\includegraphics[scale=1., bb=15 180 65 200, clip=true]
  {fig_hyperfor1.eps}}
&
sink vertex
\\
$\psi_i$
&
\raisebox{-8pt}{%
\setlength{\unitlength}{10pt}
\begin{picture}(0,0)
\put(1.6,1){$i$}
\end{picture}%
\includegraphics[scale=1., bb=15 160 65 180, clip=true]
  {fig_hyperfor1.eps}}
&
source vertex
\\
\hline
\multicolumn{3}{c}{\quad} \\[-2mm]
\hline
\multicolumn{3}{|c|}{Factors coming from 
$\prod \widehat{\mathcal{S}}_A$%
\raisebox{-6pt}{\rule{0pt}{19pt}}
}
\\
\hline
$\tau_A$
&
\raisebox{-20pt}{%
\setlength{\unitlength}{10pt}
\begin{picture}(0,0)
\put(0.2,3.2){$A$}
\end{picture}%
\includegraphics[scale=1., bb=15 105 65 155, clip=true]
  {fig_hyperfor1.eps}}
&
root hyperedge
\\
$\tau_{A \setminus i}$
&
\raisebox{-20pt}{%
\setlength{\unitlength}{10pt}
\begin{picture}(0,0)
\put(0.2,3.2){$A$}
\put(4.45,3.6){$i$}
\end{picture}%
\includegraphics[scale=1., bb=15 55 65 105, clip=true]
  {fig_hyperfor1.eps}}
&
pointing hyperedge
\\
$\psi_i \psibar_j \tau_{A \setminus \{i,j\} }$
&
\raisebox{-20pt}{%
\setlength{\unitlength}{10pt}
\begin{picture}(0,0)
\put(0.2,3.2){$A$}
\put(4.45,3.6){$j$}
\put(1.2,0.6){$i$}
\end{picture}%
\includegraphics[scale=1., bb=15 5 65 55, clip=true]
  {fig_hyperfor1.eps}}
&
dashed hyperedge
\\
\hline
\end{tabular}
\end{center}
\vspace*{-2mm}
\caption{
   Graphical representation of the various factors
   in the expansion \protect\reff{eq.A1b2}.
}
\label{table_graphical}
\end{table}


According to the rules of Grassmann algebra and Grassmann--Berezin integration,
we must have in total exactly one factor $\psibar_i$
and one factor $\psi_i$ for each vertex $i$.
Graphically this means that at each vertex we must have either
a single $\bullet$ or else the superposed pair $\otimes$
(please note that in many drawings we actually draw the $\circ$ and $\times$
 slightly split, in order to highlight which variable comes from which factor).
At each vertex $i$ we can have
an arbitrary number of ``pointing hyperedges'' pointing towards $i$,
as they do not carry any fermionic field:
\[
\setlength{\unitlength}{75pt}
\begin{picture}(2,1.2)(0,0)
\put(0,0){\includegraphics[scale=1.5, bb=260 260 360 320, clip=true]
  {fig_hyperfor1.eps}}
\put(0.975,0.15){$i$}
\end{picture}
\]
Aside from pointing hyperedges, we must be, at each vertex $i$,
in one of the following situations (Figure~\ref{fig3}):
%
%
\begin{enumerate}
   \item If $i \in V'$ or $i \in {\sf I} \cap {\sf J}$
      [resp.~cases (a) and (b) in the figure],
      the quantity $\scro_{I \cup V',J \cup V'}$ provides already
      a factor $\psibar_i \psi_i$;  therefore, no other factors
      of $\psibar_i$ or $\psi_i$ should come from the expansion
      of $\prod \widehat{\mathcal{S}}_A$.
   \item If $i \in {\sf I} \setminus {\sf J}$,
      the quantity $\scro_{I \cup V',J \cup V'}$ provides already
      a factor $\psibar_i$;  therefore, the expansion
      of $\prod \widehat{\mathcal{S}}_A$ must provide $\psi_i$,
      i.e.\ we must have one dashed hyperedge pointing from $i$.
   \item If $i \in {\sf J} \setminus {\sf I}$,
      the quantity $\scro_{I \cup V',J \cup V'}$ provides already
      a factor $\psi_i$;  therefore, the expansion
      of $\prod \widehat{\mathcal{S}}_A$ must provide $\psibar_i$,
      i.e.\ we must have one dashed hyperedge pointing towards $i$.
   \item If $i \notin {\sf I} \cup {\sf J} \cup V'$, then
      the quantity $\scro_{I \cup V',J \cup V'}$ provides
      neither $\psibar_i$ nor $\psi_i$;  therefore, the expansion
      of $\prod \widehat{\mathcal{S}}_A$ must provide
      both $\psibar_i$ and $\psi_i$,
      so that at $i$ we must have one of the following configurations:
        \begin{itemize}
            \item[a)] a non-pointed vertex of a pointing hyperedge;
            \item[b)] a vertex of a dashed hyperedge that is neither
                 of the two endpoints of the dashed arrow;
            \item[c)] a vertex of a root hyperedge;
            \item[d)] two dashed hyperedges, one with the arrow incoming,
                 one outgoing.
        \end{itemize}
\end{enumerate}

\begin{figure}[tp]
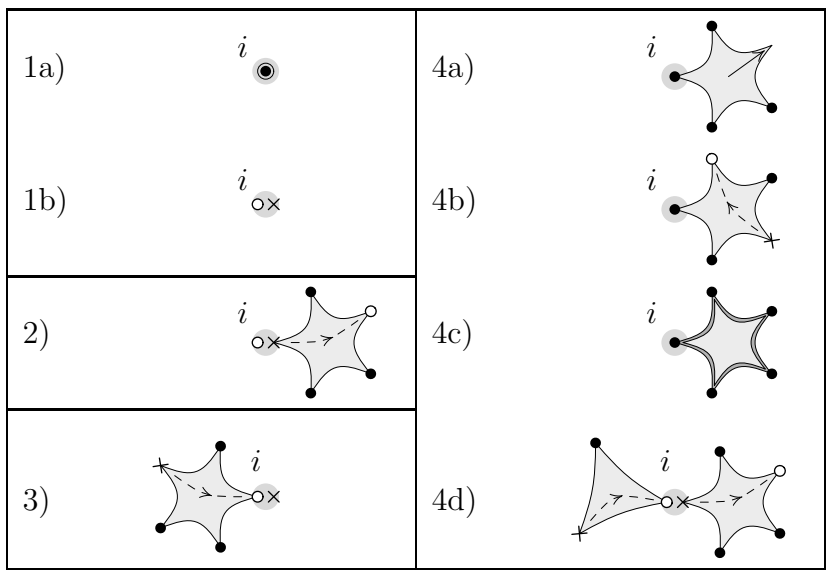

\begin{center}
\begin{tabular}{|lc|lc|}
\hline
1a) &
\setlength{\unitlength}{50pt}
\begin{picture}(0,0)
\put(0.95,0.15){$i$}
\end{picture}
\raisebox{-8pt}{
\includegraphics[scale=1., bb=70 200 170 220, clip=true]
  {fig_hyperfor1.eps}}
&
4a) &
\setlength{\unitlength}{50pt}
\begin{picture}(0,0)
\put(0.95,0.15){$i$}
\end{picture}
\raisebox{-25pt}{
\includegraphics[scale=1., bb=170 175 270 225, clip=true]
  {fig_hyperfor1.eps}}
\\
1b) &
\setlength{\unitlength}{50pt}
\begin{picture}(0,0)
\put(0.95,0.15){$i$}
\end{picture}
\raisebox{-8pt}{
\includegraphics[scale=1., bb=70 180 170 200, clip=true]
  {fig_hyperfor1.eps}}
&
4b) &
\setlength{\unitlength}{50pt}
\begin{picture}(0,0)
\put(0.95,0.15){$i$}
\end{picture}
\raisebox{-25pt}{
\includegraphics[scale=1., bb=170 125 270 175, clip=true]
  {fig_hyperfor1.eps}}
\\
\cline{1-2}
2) &
\setlength{\unitlength}{50pt}
\begin{picture}(0,0)
\put(0.95,0.15){$i$}
\end{picture}
\raisebox{-25pt}{
\includegraphics[scale=1., bb=70 125 170 175, clip=true]
  {fig_hyperfor1.eps}}
&
4c) &
\setlength{\unitlength}{50pt}
\begin{picture}(0,0)
\put(0.95,0.15){$i$}
\end{picture}
\raisebox{-25pt}{
\includegraphics[scale=1., bb=170 75 270 125, clip=true]
  {fig_hyperfor1.eps}}
\\
\cline{1-2}
3) &
\setlength{\unitlength}{50pt}
\begin{picture}(0,0)
\put(1.05,0.3){$i$}
\end{picture}
\raisebox{-20pt}{
\includegraphics[scale=1., bb=70 75 170 125, clip=true]
  {fig_hyperfor1.eps}}
&
4d) &
\setlength{\unitlength}{50pt}
\begin{picture}(0,0)
\put(1.05,0.3){$i$}
\end{picture}
\raisebox{-22pt}{
\includegraphics[scale=1., bb=170 15 270 75, clip=true]
  {fig_hyperfor1.eps}}
\\
\hline
\end{tabular}
\end{center}
\caption{
   Possible ways of saturating the Grassmann fields on vertex $i$
   (indicated by the small gray disk).
}
\label{fig3}
\end{figure}

Having given the local description of the possible configurations
at each vertex $i$, let us now describe the possible global configurations.
Note first that at each vertex we can have at most two incident dashed arrows,
and if there are two such arrows then they must have opposite orientations.
As a consequence, we see that dashed arrows must either form cycles,
or else form open paths connecting
a source vertex of ${\sf I} \setminus {\sf J}$
to a sink vertex of ${\sf J} \setminus {\sf I}$.
Let us use the term {\em root structures}\/ to denote root vertices,
root hyperedges, cycles of dashed hyperedges,
and open paths of dashed hyperedges.

As for the solid arrows in the pointing hyperedges, the
reasoning is as follows:  If a pointing hyperedge $A$ points towards
$i$, then either $i$ is part of a root structure as described above,
or else it is a non-pointed vertex of another pointing hyperedge $\varphi(A)$.
We can follow this map iteratively, i.e.~go to $\varphi(\varphi(A))$,
and so on:
\[
\setlength{\unitlength}{10pt}
\begin{picture}(0,0)
\put(0.2,3.2){$A$}
\put(6.,1.6){$\phi(A)$}
\put(12.8,3){$\phi(\phi(A))$}
\put(14,1){$\cdots$}
\end{picture}
\includegraphics[scale=1., bb=10 238 170 288, clip=true]
  {fig_hyperfor1.eps}
\]
Because of the finiteness of the graph,
either we ultimately reach a root structure, or we enter a cycle.
Cycles of the ``dynamics'' induced by $\varphi$ correspond
to cycles of the pointing hyperedges.
We now also include such cycles of pointing hyperedges
as a fifth type of root structure
(see Figure~\ref{fig4} for the complete list of root structures).

\begin{figure}[t]
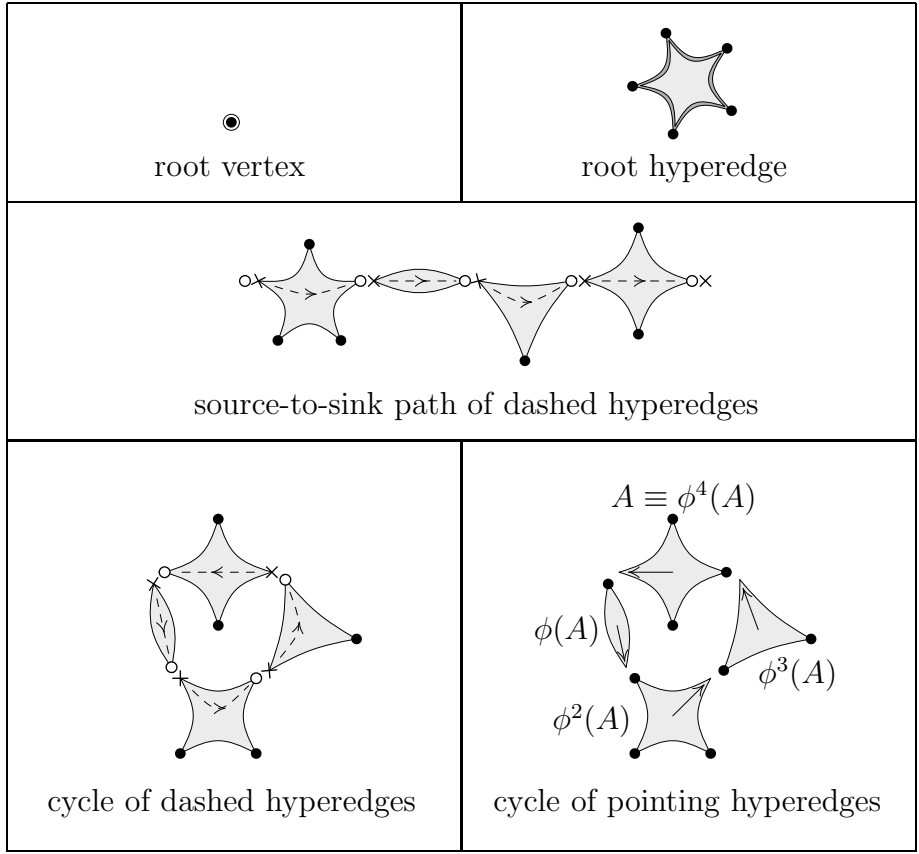

\begin{center}
\setlength{\unitlength}{10pt}
\begin{picture}(32,31)
\put(-1.5,-1.5){\line(1,0){34}}
\put(-1.5,14){\line(1,0){34}}
\put(-1.5,23){\line(1,0){34}}
\put(-1.5,30.5){\line(1,0){34}}
\put(-1.5,-1.5){\line(0,1){32}}
\put(15.5,-1.5){\line(0,1){15.5}}
\put(15.5,23){\line(0,1){7.5}}
\put(32.5,-1.5){\line(0,1){32}}
\put(6,25.5){
\includegraphics[scale=1., bb=35 205 45 215, clip=true]
  {fig_hyperfor1.eps}}
\put(4,24){root vertex}
\put(21,25){
\includegraphics[scale=1., bb=15 105 65 155, clip=true]
  {fig_hyperfor1.eps}}
\put(20,24){root hyperedge}
\put(5.8,16.5){
\includegraphics[scale=1., bb=15 295 205 355, clip=true]
  {fig_hyperfor1.eps}}
\put(5.5,15){source-to-sink path of dashed hyperedges}
\put(3,2){
\includegraphics[scale=1., bb=300 30 385 125, clip=true]
  {fig_hyperfor1.eps}}
\put(0,0){cycle of dashed hyperedges}
\put(20,2){
\includegraphics[scale=1., bb=300 130 385 225, clip=true]
  {fig_hyperfor1.eps}}
\put(16.7,0){cycle of pointing hyperedges}
\put(18.2,6.5){$\phi(A)$}
\put(18.9,3.1){$\phi^2(A)$}
\put(26.6,4.85){$\phi^3(A)$}
\put(21.1,11.5){$A \equiv \phi^4(A)$}
\end{picture}
\end{center}
\caption{
   The five kinds of root structures.
}
\label{fig4}
\end{figure}

All the rest is composed of pointing hyperedges,
which form directed arborescences,
rooted on the vertices of the root structures.
In conclusion, therefore, the most general configuration
consists of a bunch of disjoint root structures,
and a set of directed arborescences (possibly reduced to a single vertex)
rooted at its vertices, such that the whole is
a spanning subhypergraph $H$ of $G$.

As each root structure is either a single vertex, a single hyperedge,
a (hyper-)path or a (hyper-)cycle,
we see that each connected component of $H$ is either
a (hyper-)tree or a (hyper-)unicyclic.
Furthermore, all vertices in ${\sf I} \cup {\sf J}$
are in the tree components,
and each tree contains either
one vertex from ${\sf I}$ and one from ${\sf J}$ (possibly coincident)
or else no vertices at all from ${\sf I} \cup {\sf J}$.

We still need to understand the weights associated to the allowed
configurations.
Clearly, we have a factor $w_{A;i}$ per pointing hyperedge in the
arborescence. Root vertices coming from $V'$ have factors $t_i$, and
root hyperedges have factors $\widehat{w}_A^*$. 
Cycles $\gamma = (i_0,A_1, i_1, A_2, ...., i_{\ell} = i_0)$
of the dynamics of $\varphi$ (\emph{bosonic cycles}) have a weight
$w_{A_1; i_1} \cdots w_{A_{\ell}; i_{\ell}}$.
All the foregoing objects contain Grassmann variables
only in the combination $\psibar_i \psi_i$,
and hence are commutative.
Finally, we must consider the dashed hyperedges,
which contain ``unpaired fermions'' $\psi_i$ and $\psibar_j$,
and hence will give rise to signs coming from anticommutativity.
Let us first consider the dashed cycles
$\gamma = (i_0,A_1, i_1, A_2, ...., i_{\ell} = i_0)$,
and note what happens when reordering the fermionic fields:
\begin{eqnarray}
& &
(w_{A_1; i_{\ell} i_1} \psi_{i_{\ell}} \psibar_{i_1})
(w_{A_2; i_1 i_2} \psi_{i_1} \psibar_{i_2})
\cdots
(w_{A_{\ell}; i_{\ell-1} i_{\ell}} \psi_{i_{\ell-1}} \psibar_{i_{\ell}})
   \nonumber \\
& & \qquad =\;
- w_{A_1; i_{\ell} i_1} w_{A_2; i_1 i_2} \cdots
    w_{A_{\ell}; i_{\ell-1} i_{\ell}} 
\,
\psibar_{i_1} \psi_{i_1}
\cdots
\psibar_{i_{\ell}} \psi_{i_{\ell}}
\end{eqnarray}
because $\psi_{i_{\ell}}$ had to pass through $2\ell-1$ fermionic fields
to reach its final location.
This is pretty much the result one would have expected, \emph{but}\/ we
have an overall minus sign, irrespective of the length of the cycle
(or its parity), which is in a sense ``non-local'', due to the
fermionic nature of the fields $\psi$ and $\psibar$.
For this reason we call a dashed cycle a \emph{fermionic cycle}\/.

A similar mechanism arises for the open paths of dashed hyperedges
$\gamma = (i_0,A_1, i_1,$ $A_2, \ldots, i_{\ell})$,
where $i_0$ is the source vertex and $i_{\ell}$ is the sink vertex.
Here the weight
$w_{A_1;i_0 i_1} w_{A_2; i_1 i_2} \cdots w_{A_\ell; i_{\ell-1} i_\ell}$
multiplies the monomial
$\psi_{i_0} \psibar_{i_1} \psi_{i_1} \psibar_{i_2} \psi_{i_2}
 \cdots \psibar_{i_{\ell-1}} \psi_{i_{\ell-1}} \psibar_{i_\ell}$,
in which the only unpaired fermions are $\psi_{i_0}$ and $\psibar_{i_\ell}$.
in this order.  Now the monomials for the open paths must be
multiplied by $\scro_{I,J}$, and each source (resp.\ sink) vertex
from an open path must correspond to a vertex of ${\sf I}$ (resp.\ ${\sf J}$).
This pairing thus induces a permutation of $\{1,\ldots,k\}$,
where $k = |{\sf I}| = |{\sf J}|$:  namely, $i_r$ is connected by an open path
to $j_{\pi(r)}$.  We then have
\begin{equation}
   \Biggl( \prod_{r=1}^k \psibar_{i_r} \psi_{j_r} \Biggr)
   \Biggl( \prod_{r=1}^k \psi_{i_r} \psibar_{j_{\pi(r)}} \Biggr)
   \;,
\end{equation}
where the first product is $\scro_{I,J}$ and the second product
comes from the open paths.  This can easily be rewritten as
\begin{subeqnarray}
   \prod_{r=1}^k \psibar_{i_r} \psi_{j_r} \psi_{i_r} \psibar_{j_{\pi(r)}}
   & = &
   \prod_{r=1}^k \psibar_{i_r} \psi_{i_r} \psibar_{j_{\pi(r)}} \psi_{j_r} 
        \\[1mm]
   & = &
   \Biggl( \prod_{r=1}^k \psibar_{i_r} \psi_{i_r} \Biggr)
   \Biggl( \prod_{r=1}^k \psibar_{j_{\pi(r)}} \psi_{j_r} \Biggr)
        \\[1mm]
   & = &
   \sgn(\pi) \,
   \Biggl( \prod_{r=1}^k \psibar_{i_r} \psi_{i_r} \Biggr)
   \Biggl( \prod_{r=1}^k \psibar_{j_r} \psi_{j_r} \Biggr)
       \;.
\end{subeqnarray}

Putting everything together, we see that the Grassmann integral \reff{eq.A1}
can be represented as a sum over rooted oriented spanning subhypergraphs
$\vec{H}$ of $G$, as follows:
\begin{itemize}
   \item  Each connected component of $H$ (the unoriented subhypergraph
      corresponding to $\vec{H}$) is either a (hyper-)tree or a
      (hyper-)unicyclic.
   \item  Each (hyper-)tree component contains either one vertex from
      ${\sf I}$ (the {\em source vertex}\/) and one from ${\sf J}$
      (the {\em sink vertex}\/, which is allowed to coincide with the
      source vertex), or else no vertex from ${\sf I} \cup {\sf J}$.
      In the latter case, we choose either one vertex of the component
      to be the {\em root vertex}\/, or else one hyperedge of the
      component to be the {\em root hyperedge}\/.
   \item Each unicyclic component contains no vertex from
      ${\sf I} \cup {\sf J}$.  As a unicyclic, it necessarily
      has the form of a single (hyper-)cycle together with
      (hyper-)trees (possibly reduced to a single vertex)
      rooted at the vertices of the (hyper-)cycle.
   \item Each hyperedge other than a root hyperedge is {\em oriented}\/
      by designating a vertex $i(A) \in A$ as the {\em outgoing vertex}\/.
      These orientations must satisfy following rules:
         \begin{itemize}
             \item[(i)]  each (hyper-)tree component is directed towards
                 the sink vertex, root vertex or root hyperedge,
             \item[(ii)]  each (hyper-)tree belonging to a unicyclic
                 component is oriented towards the cycle, and
             \item[(iii)]  the (hyper-)cycle of each unicyclic component
                 is oriented consistently.
         \end{itemize}
      Thus, in each (hyper-)tree component the orientations are fixed
      uniquely, while in each unicyclic component we sum over the two
      consistent orientations of the cycle.
\end{itemize}
The {\em weight}\/ of a configuration $\vec{H}$ is the product of
the weights of its connected components, which are in turn defined
as the product of the following factors:
\begin{itemize}
   \item  Each root vertex $i$ gets a factor $t_i$.
   \item  Each root hyperedge $A$ gets a factor $\widehat{w}_A^*$.
   \item  Each hyperedge $A$ belonging to the (unique) path from
      a source vertex to a sink vertex gets a factor $w_{A;ij}$,
      where $j$ is the outgoing vertex of $A$ and $i$ is the
      outgoing vertex of the preceding hyperedge along the path
      (or the source vertex if $A$ is the first hyperedge of the path).
   \item  Each hyperedge $A$ that does not belong to a source-sink path
      or to a cycle gets a factor $w_{A;i(A)}$
      [recall that $i(A)$ is the outgoing vertex of $A$].
   \item  Each oriented cycle $(i_0,A_1, i_1, A_2, ...., i_{\ell} = i_0)$
      gets a weight
      \begin{equation}
         \prod_{\alpha=1}^\ell w_{A_\alpha;i_\alpha} \,-\,
         \prod_{\alpha=1}^\ell w_{A_\alpha;i_{\alpha-1} i_\alpha}
         \;.
      \end{equation}
   \item  There is an overall factor $\sgn(\pi)$.
\end{itemize}

\subsection{Special cases}

The contribution from unicyclic components cancels out whenever
$\prod_{\alpha=1}^\ell w_{A_\alpha;i_\alpha} =
 \prod_{\alpha=1}^\ell w_{A_\alpha;i_{\alpha-1} i_\alpha}$
for every oriented cycle
$(i_0,A_1, i_1, A_2, ...., i_{\ell} = i_0)$.
In particular, this happens if $w_{A;ij} = w_{A;j}$
for all $A$ and all $i,j \in A$.
More generally, it happens if $w_{A;ij} = w_{A;j} \exp(\phi_{A;ij})$
where $\phi$ has ``zero circulation'' in the sense that
$\sum_{\alpha=1}^\ell \phi_{A_\alpha;i_{\alpha-1} i_\alpha} = 0$
for every oriented cycle
$(i_0,A_1, i_1, A_2, ...., i_{\ell} = i_0)$.
Physically, $\phi$ can be thought of as a kind of ``gauge field'
to which the fermions $\psi,\psibar$ are coupled;
the zero-circulation condition means that $\phi$ is
gauge-equivalent to zero.
Note, finally, that if $w_{A;i} w_{A;j} = w_{A;ij} w_{A;ji}$
for all $i,j \in A$, then $\widehat{w}_A^* = w_A^*$.

At the other extreme, if we take all $t_i = 0$,
all $\widehat{w}_A^* = 0$ and $I = J = \emptyset$,
then all tree components disappear, and we are left with only unicyclics.

In certain ``symmetric'' circumstances,
we can combine the contributions from tree components
having the same set of (unoriented) hyperedges but different roots,
and obtain reasonably simple expressions.
In particular, suppose that the weights $w_{A;i}$ are independent of $i$
(let us call them simply $w_A$),
and consider a tree component $T$ that does not contain
any vertices of ${\sf I} \cup {\sf J}$.
Then we can sum over all choices of root vertex or root hyperedge,
and obtain the weight
\begin{equation}
   \Biggl( \prod_{A \in E(T)} w_A \Biggr)
   \Biggl( \sum_{i \in V(T)} t_i \,+\, 
           \sum_{A \in E(T)} \frac{\widehat{w}_A^*}{w_A}  \Biggr)
   \;.
 \label{eq.special.sym}
\end{equation}
A further simplification occurs in two cases:
\begin{itemize}
   \item If all $t_i = t$ and all $\widehat{w}_A^* = 0$,
      then the second factor in \reff{eq.special.sym} becomes
      simply $t |V(T)|$:  we obtain forests of
      {\em vertex-weighted trees}\/.
   \item If all $t_i = t$ and $\widehat{w}_A^* = t(1-|A|) w_A$
      for all $A$, then the second factor in \reff{eq.special.sym}
      becomes simply $t$
      (by virtue of Proposition~\ref{prop.eulerineq.hypergraphs})
      and we obtain {\em unrooted}\/ forests.
\end{itemize}
Recall, finally, that if we also take $w_{A;ij} = w_A$
for all $A$ and all $i,j \in A$, then the unicyclic components cancel
and $\widehat{w}_A^* = w_A^*$, so that \reff{eq.special.sym}
reduces to \reff{eq.thm.Zgrass.gen}.

It is instructive to consider the special case in which $G$ is an
ordinary graph, i.e.\ each hyperedge $A \in E$ is of cardinality 2.
If we further take all $w_A^* = 0$,
then the quantity in the exponential of the functional integral \reff{eq.A1}
is a quadratic form
$\mathcal{S}(\psi,\psibar) + \sum\limits_{i \in V} t_i \psibar_i \psi_i
 = \psibar M \psi$,
with matrix
\begin{equation}
   M_{ij}  \;=\;
   \begin{cases}
      t_i + \sum\limits_{k \neq i} w_{\{i,k\};k}  & \hbox{if }  i = j     \\
      -w_{\{i,j\};ji}                             & \hbox{if }  i \neq j
   \end{cases}
\end{equation}
Our result for $I = J = \emptyset$
then corresponds to the ``two-matrix matrix-tree theorem''
of Moon \cite[Theorem~2.1]{Moon}
with $r_{ik} = w_{\{i,k\};k}$ for $i \neq k$, $r_{ii} = 0$,
$s_{ij} = w_{\{i,j\};ij}$ for $i \neq j$ and $s_{ii} = -t_i$.\footnote{
   There is a slight notational difference between us and Moon \cite{Moon}:
   he has the bosonic and fermionic cycles going in the same direction,
   while we have them going in opposite directions.
   But this does not matter, because $\det(M) = \det(M^{\rm T})$.
   Our ``transposed'' notation was chosen in order to
   make more natural the definitions of correlation functions
   in Section~\ref{sec:corrfn}.
}

\section*{Acknowledgments}

We wish to thank Andrea Bedini for helpful discussions
on many aspects of this formalism,
and Jean Bricmont and Antti Kupiainen
for helpful discussions concerning
\reff{eq.2pointfn}--\reff{eq.2pointfn.special}.
We are also grateful to an anonymous referee
for suggestions that greatly improved
Sections~\ref{sec:grass} and \ref{sec:osp}.

This work was supported in part by U.S.~National Science Foundation
grant PHY--0424082.
One of us (Sportiello) is grateful to New York University,
Oxford University and Universit\'e de Paris-Sud (Orsay) for kind hospitality.
We also wish to thank LPTHE--Jussieu for hospitality
while this article was being finished.



\end{document}